\documentclass[useAMS,usenatbib]{mn2e}

\usepackage{Times}
\usepackage{graphicx}

\title[A universal stellar-mass and size relation]{A universal stellar-mass and size relation of galaxies in GOODS-N region}
\author[T. Ichikawa et al.]
{Takashi Ichikawa,$^{1}$\thanks{E-mail:
ichikawa@astr.tohoku.ac.jp}
Masaru Kajisawa,$^{2}$,
and 
Mohammad Akhlaghi$^{1}$\\
$^{1}$Astronomical Institute, Tohoku University, Aramaki, Aoba, Sendai 980-8578, Japan\\
$^{2}$Research Center for Space and Cosmic Evolution, Ehime University, Bunkyo-cho 2-5, Matsuyama 790-8577, Japan
}
\begin{document}

\date{Accepted. Received; in original form }

\pagerange{\pageref{firstpage}--\pageref{lastpage}} \pubyear{}

\maketitle

\label{firstpage}

\begin{abstract}
We present scaling relations between stellar-mass ($M_*$) and the size of galaxies  at $0.3<z<3$ for 
half- ($R_{50}$) and 90 percent-light ($R_{90}$) radii, using a deep $K$-band selected catalogue 
taken with the Subaru Telescope and MOIRCS in the GOODS-North region. 
The logarithmic slope $R \propto M_*^{0.1\sim0.2}$ is independent of redshift in a wide mass range of 
$M_*\sim 10^{8}-10^{11}$ M$_\odot$, irrespective of galaxy populations (star-forming, quiescent). 
The offset change is $\la 50$ percent.
Provided that optical light in the rest frame traces the stellar mass of galaxies,
the universal relation demonstrates that the stellar mass was built up in galaxies over their cosmic 
histories  in a similar manner on average irrelevant to galaxy mass.
The small offset in each stellar mass bin from the universal relation shows weak size evolution at a given mass.
There is a moderate increase of 30--50 percent for $R_{50}$ and $R_{90}$ for less massive galaxies
 ($M_*<10^{10}$ M$_\odot$) from $z\sim3$ to $z\sim1$, while 
the sizes remains unchanged or slightly decrease towards $z\sim0.3$.
For massive galaxies ($M_*\ga10^{11}$ M$_\odot$), the evolution is $\sim70-80$ \% increase in $R_{90}$ 
from $z\sim3$ to $z\sim0.3$, though that in $R_{50}$ is weaker.
The evolution of compactness factor, $R_{50}/R_{90}$, which becomes smaller at lower redshift,
 is suggestive of minor merging effect in the outer envelope of massive galaxies.

\end{abstract}


\section{INTRODUCTION}
The scale size of galaxies is one of the fundamental parameters to elucidate the history of galaxy 
formation and evolution.
The change of size and stellar-mass relations over cosmic time would pose strong constraints on models 
of galaxy evolution.
The observational relations between galaxy size and stellar mass have been  
studied in the local universe, based on the Sloan Digital Sky Survey (Shen et al. 2003; Bernardi et al. 2011).
Using rest-frame optical bands,  which presumably trace  the distribution of stellar mass in galaxies,
many studies have investigated galaxy sizes at higher redshift as a function of stellar mass for massive 
galaxies ($M_*\ga 10^{10}$ M$_\odot$).
For example, the relations for galaxies at $0.2<z<1$ were studied for late-type 
galaxies (Barden et al. 2005) and  for early-type galaxies (McIntosh et al. 2005; Trujillo, Ferreras, \& de la Rosa 2011).
Damjanov et al. (2009) and Mancini et al. (2010) gave size-mass relations of massive galaxies ($M_* > 10^{10.5}$ M$_\odot$) 
at $z\sim1.5$.
Williams et al. (2010) studied the relation with large samples of galaxies at $z\la2$.
For higher redshifts to $z\sim3$, size-mass relations have been obtained for galaxies with $M_* \ga 10^{10}$ M$_\odot$ (e.g.,
Franx et al. 2008; Nagy et al. 2011; Cassata et al. 2011).

Many studies have corroborated that massive galaxies at high redshifts were much smaller 
than local galaxies with comparable mass (e.g., Daddi et al. 2005; 
Trujillo et al. 2006, 2007; Toft et al. 2007; Zirm et al. 2007; Cimatti et al. 2008; Buitrago et al. 2008; 
van Dokkum et al. 2008, 2009, 2010; Akiyama et al. 2008; Damjanov et al. 2009; Carrasco et al. 2010; Cassata et al. 2010;
Szomoru et al. 2010; van der Wel 2011; Cassata et al. 2011).
At a fixed stellar mass, spheroidal galaxies were significantly more compact at high redshift 
and evolved with rapid increase of the effective radius by a factor $\sim$4 or even larger from $z\sim 2$ 
(e.g., Buitrago et al. 2008; Carrasco et al. 2010)
and by a factor $\sim$2 from $z\sim1$ (e.g., van der Wel et al. 2008; Trujillo et al. 2011).
The finding of compact massive galaxies with a high stellar velocity dispersion also supports
their existence (van Dokkum 2009; van de Sande 2011).
It contrasts with the absence of such compact massive galaxies in the local universe, though several candidates 
have been found at $z\sim0.5$ (Stockton et al. 2010) and in the local universe (Trujillo et al. 2009; Valentinuzzi et al. 2010).
The findings demonstrate that massive galaxies have increased their size dramatically since $z\sim 3$ 
in a different manner from the evolution of less massive galaxies.

However, other studies have reached contradictory conclusions.
There is significant disagreement between the results of different studies.
Barden et al. (2005) found weak or no evolution in the relation between stellar mass and effective 
disc size for galaxies with $M_*>10^{10}$ M$_\odot$ since $z\sim1$.
For early-type galaxies at $z\sim1$, McIntosh et al. (2005) showed that luminosity-size and stellar mass 
size relations evolve in a manner that is consistent with a completely passive evolution of
the red early-type galaxy population.
It is also shown that not all high-redshift early-type galaxies were compact and underwent dramatic size evolution 
 (e.g., Toft et al. 2007; Zirm et al. 2007; Saracco et al. 2009, 2011; Mancini et al. 2010; Stott et al, 2011).
From the study of surface brightness in rest-frame $V$ and $z$ bands at $z\la3$, Ichikawa et al. (2010) gave another evidence 
for no conspicuous evolution in galaxy sizes.

As many previous studies show, the size evolution of galaxies are still controversial.
Any systematic errors in the observation or analyses could bias results towards such a significant evolution 
(e.g., Mancini et al. 2010; Hopkins et a. 2009b; Bouwens et al. 2004).
The origin of the discrepancy could be ascribed to redshift effects that more distant galaxies look 
more compact due to the difficulty of measuring envelopes at low surface density.
The light from the outer portion of high-redshift galaxies is apt to be hidden in noise for low $S/N$ 
observations.
As the consequence, effective radii and total luminosity (or stellar mass) would be underestimated.
On the other hand, very deep observations (e.g., Szomoru et al. 2010; Cassata et al. 2010; Law et al. 2012) or stacking 
methods to enhance the faint envelope of galaxies (e.g., Zirm et al. 2007; van Dokkum et al. 2008, 2010; van der Wel et al. 2008) 
have claimed that it is not the case.

How significantly have the sizes of less-massive normal galaxies evolved from the early universe to the current epoch?
In this context, we look into the evolution of stellar-mass and size scaling relations 
on the basis of half- and 90 percent-light encircles, focusing on less massive galaxies ($M_*<10^{11}$ M$_\odot$) 
at $0.3<z<3$, using a deep $K$-selected galaxy catalogue.
We infer that the outer radius is more influenced by merging effect than star formation or central activity. 
In \S2, we describe the catalogue we used, which is among the deepest in the $K$ band to date.
The depth is crucial for studying galaxies of low-surface brightness or galaxies which are dimmed due to the cosmological expansion 
at high redshift to measure the radius at faint outskirt of galaxies.
The data analysis and the result for the size and stellar-mass relation are 
detailed in \S3 and  \S4. The results are discussed in \S5.
Ichikawa et al. (2010) studied the evolution of surface brightness of galaxies at $z\la3$ in 
rest-frame $V$ and $z$ bands with the same data as the present study.
We will discuss the consistency of the present result with their study.

Throughout this paper, we assume $\Omega_m =0.3$, 
$\Omega_\Lambda =0.7$, and $H_{\mathrm{0}} =70$ km s$^{-1}$ Mpc$^{-1}$.
We use the AB magnitude system (Oke \& Gunn 1983; Fukugita et al. 1996). 

\section[]{DATA}

We use the $K$-band selected catalogue of the MOIRCS Deep Survey (MODS) in the GOODS--North region 
(Kajisawa et al. 2009, hereafter K09; Kajisawa et al. 2011, K11),  which are based on our imaging observations 
in $JHKs$ bands with MOIRCS  (Suzuki et al. 2008) and archived data.
Four MOIRCS pointings cover 70 percent of the GOODS--North region (103 arcmin$^2$, hereafter referred 
as `wide' field). 
One of the four pointings, which includes HDF-N (Williams et al. 1996), is  
the ultra-deep field of MODS (28 arcmin$^2$,  `deep' field).
As the accuracy of background subtraction was highly demanded for the study of faint 
end of galaxies, the background was scrutinized and carefully subtracted (see K11 for the details).
The surface brightness limit was extensively examined in K11.
The typical $1\sigma$ surface brightness fluctuations in one arcsec diameter were found to be $\sim$27 mag 
arcsec$^{-2}$ and $\sim$27.5 mag arcsec$^{-2}$ in $K$ band for the wide and deep fields respectively.
These are $1\sim2$ mag deeper than those of previous studies in $K$ band.
The depth is crucial for the study of low surface brightness features at the outskirts of galaxies because of 
the strong dependence of cosmological dimming of surface brightness on redshift.

For the total magnitude $m_K$ in $K$ band, we use MAG\_AUTO obtained by SExtractor (Bertin \& Arnouts 1996).
The $85\sim90$-percent completeness of the catalogue is $m_K\sim25$ in the wide field and $\sim$26 mag in the deep 
field. 
We exclude fainter galaxies from the present analysis.
The FWHMs of the final stacked images are 0.46 arcsec for the deep image
and 0.53--0.60 arcsec for the wide image. 
The numbers of galaxies are 3555 and 6063, respectively.

To obtain the stellar mass of MODS samples, K09 performed SED fitting of the 
multiband photometry ($UBVizJHK$, 3.6 $\mu$m, 4.5 $\mu$m, and 5.8 $\mu$m) with population 
synthesis models.
We adopt the results with GALAXEV templates (Bruzual \& Charlot 2003) and the Salpeter initial mass
function (see K09 for more details).
The stellar masses ($M_*$) are obtained from the best-fit stellar mass-to-luminosity ratio in $K$ band
and scaled with the $K$-band flux. 
A detailed comparison between different mass estimators is given in K09.
In the present analysis, the near-infrared data (3.6$\mu$m) of Spitzer/IRAC 
are available for most of the sample galaxies (96 percent), so that SED fitting is reasonably 
reliable for the photometry at rest-$V$ ($\lambda_\mathrm{eff} = 0.55 \mu$m) out to redshift $\sim$3.
We assume that the size of the stellar system in galaxies is represented by the size measured on $K$-band
images, which are the rest-frame optical to near-infrared wavelengths at $z\la3$.
The difference of the sizes between optical and infrared light is discussed in \S4.

In the present catalogue, 209 galaxies are identified as X-ray sources.
Among them, 61 sources are massive galaxies emitting hard X-rays
with $M_*>10^{10.5}$ M$_\odot$ at $2<z<4$ (Yamada et al. 2009).
One would expect galaxies with AGN emission to have smaller half-light radius.
Considering possible effects on the size and mass estimates, we discard X-ray sources in what
follows. 

\section[]{ANALYSES}

\subsection{Galaxy sizes}

As the definition of galaxy size, most previous studies have used the scale length ($r_e$),
 obtained by fitting the S\'ersic profile (S\'ersic 1968) to galaxy images.
The accuracy of the scale length strongly depends on image quality and depth of observations.
For bright galaxies, reliable galaxy sizes can be measured from ground-based data (Trujillo et al. 2006; 
Franx et al. 2008; Williams et al. 2010).
On the other hand, it is crucial to take an accurate measure of the profile for small objects near the resolution
limit.
In fact, Konishi et al. (2011) applied two dimensional fitting for the present sample 
to obtain S\'ersic indexes with a single component for a morphological study 
of MODS galaxies.
However, they found that reliable fitting was successful only for comparatively massive galaxies ($M_*>10^{10}$ M$_\odot$) 
at $z\sim1$ and very bright galaxies ($K<22.5$) at higher redshift.
The resolution of FWHM $=0.5 \sim 0.6$ arcsec like the present sample would not allow reliable
fitting with S\'ersic profile to fainter galaxies.
S\'ersic parameters are sometimes degenerated between $n$ and $r_e$, which depends on
the surface-brightness limit of the images (e.g., Mancini et al. 2010; Stott et al. 2011), 
so that the reliable fitting would be difficult for faint galaxies.
Moreover high-redshift galaxies exhibit a wide range of disturbed morphologies (e.g., Kajisawa \& Yamada 2001).
Therefore, the fitting for less massive galaxies at high redshift could also be strongly influenced 
by their more amorphous features in the shape.
In addition, we must take into account that we are observing a mix of galaxy morphologies.

As such, we define the half-light radius ($r_{50}$) as the radius of a circular aperture, 
which encircles half the $K$-band light emitted from galaxies.
As we here focus not just on massive (or bright) galaxies, but also on less massive galaxies
up to $z\sim3$, we prefer $r_\mathrm{50}$ to $r_e$ for the study of the structural parameters.
If a galaxy profile extends to infinity, $r_e$ encloses half the flux in the S\'ersic profile.
However, as galaxies are supposed to have an edge at several times the scale length, 
$r_e$ does not always mean half-light radius.
Therefore, in general, $r_{50}$ is smaller than $r_e$.
In the same way, we define 90 percent-light radius $r_\mathrm{90}$.
The 90-percent radius will give us more information at outer region of galaxies, where 
the size would be more influenced by merging effect.
$r_{50}$ and $r_{90}$ are obtained by SExtractor with PHOT\_FLUXFRAC=0.5 and 0.9

\begin{figure*}
\includegraphics{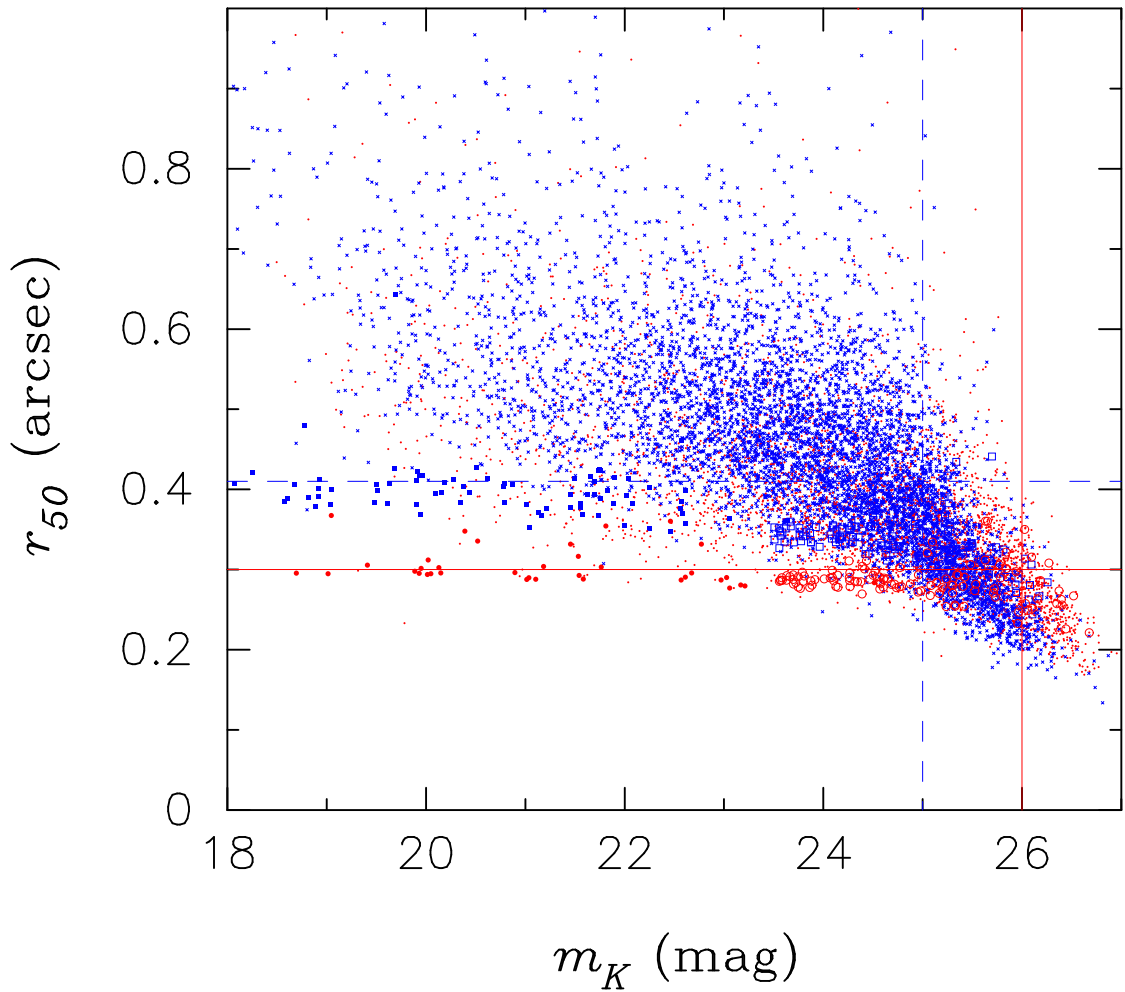}
\caption{Half-light radius ($r_\mathrm{50}$) and total magnitude ($m_K$) for
the MODS catalogue.
The dots (red) and crosses (blue) shows the samples in the deep and wide fields, respectively.
The vertical solid and dashed lines indicate the magnitude limit for each field.
Filled (blue) squares and circles (red) are spectroscopically confirmed stars.
Few stars fainter than 23.5 mag with spectroscopic data are available, so that we plot
the results of SExtractor detection of artificial stars embedded in noise images 
(open blue squares for the wide and open red circles for the deep).
The horizontal (blue) dash and (red) solid lines delineates unresolved sources. }
\label{fig-Kvsr50}
\end{figure*}

Figure \ref{fig-Kvsr50} shows $r_{50}$  versus $m_K$ for all objects in MODS. 
MODS catalogue lists many spectroscopically confirmed stars, which are also plotted in
Fig.~\ref{fig-Kvsr50}.
As few spectroscopic data were available for stars fainter than $m_K\sim23.5$  
in the present region, we examined the reliability of $r_\mathrm{50}$ using artificial stars.
The artefacts were convolved with a Moffat point spread function ($\beta=3$ and FWHM=0.5 arcsec in
the deep field and 0.6 arcsec in the wide field), and then buried in images 
as noisy as those of the deep and wide fields.
SExtractor was applied to the artefacts in the same manner as for the MODS catalogue.
The results are plotted in Fig.~\ref{fig-Kvsr50}.
The location of stars gives a clear boundary of unresolved galaxies.
We define galaxies smaller than $r_{50}=0.41$ in the wide field  $r_{50}=0.30$ in the deep field
as unresolved galaxies.

\subsection{Monte Carlo study for the sizes with mock galaxies}

\begin{figure*}
\includegraphics{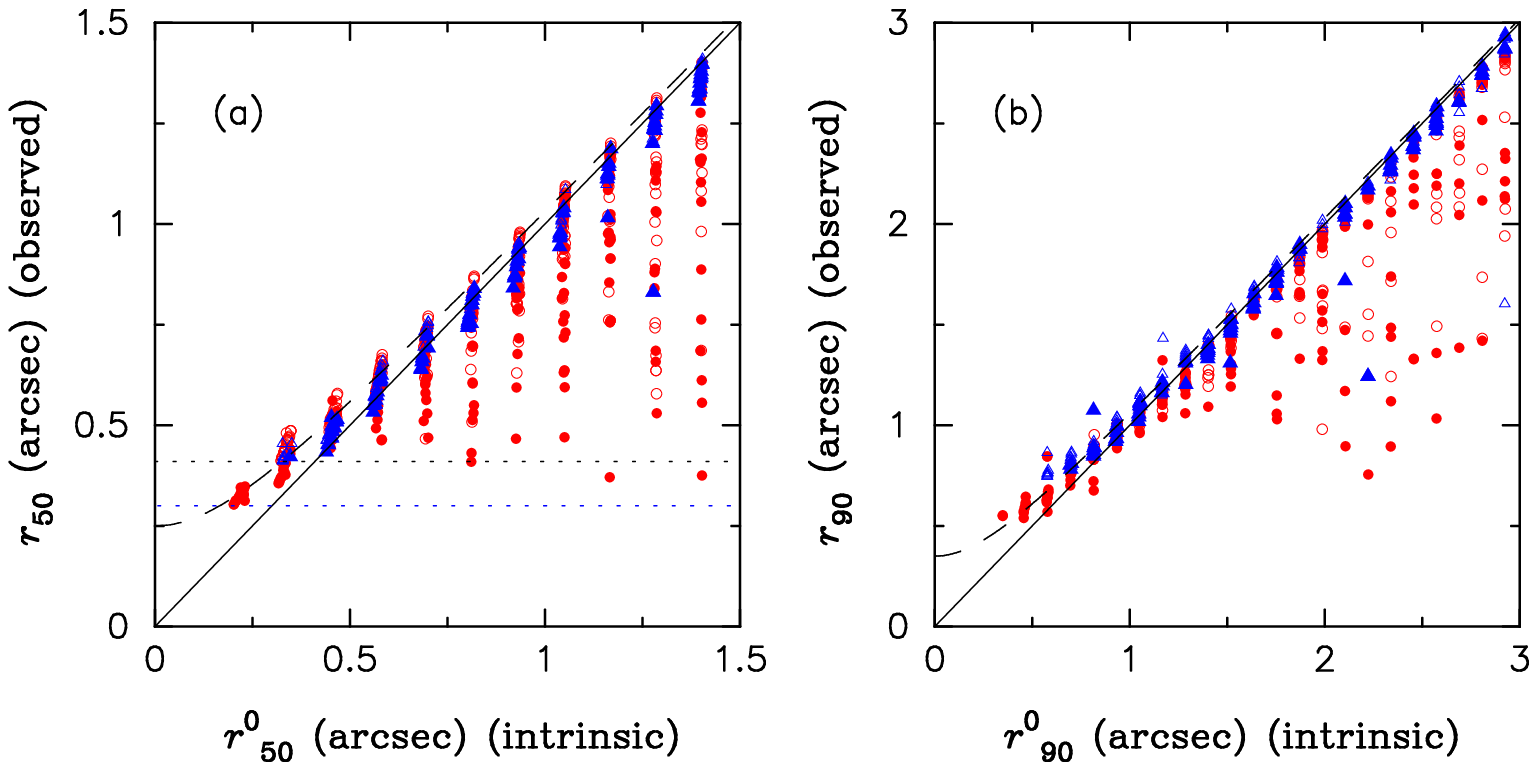}
\caption{(a) Half-light ($r_{50}$) and (b) 90 percent-light ($r_{90}$) radii of mock galaxies
embedded in the noise images.
The abscissa is the intrinsic values and the ordinate is those observed with SExtractor.
Triangles (blue) and circles (red) indicate the model galaxies with exponential and $1/4$-law profiles, respectively.
Open symbols are the sources ($m_K\leq 25$) in the noise image of the wide field, while filled ones are those 
($m_K\leq 26$) in the deep field.
The sample galaxies larger than  $r_{50}=0.41$ arcsec for the wide field and $r_{50}=0.3$ arcsec
 for the deep field are plotted above horizontal dot lines in (a).
The dashed curves show the effect of point spread function $\Delta r=0.25$ and 0.35 arcsec for 
$r_{50}$ and $r_{90}$, respectively.
}
\label{fig-mockr90r50}
\end{figure*}

We first study the reliability of the galaxy size.
Seeing and background noise strongly affect the observed $r_\mathrm{50}$ and $r_\mathrm{90}$, 
especially for faint or small galaxies.
To examine the effect on the size estimate, we generated mock galaxies with a 1/4-law or 
exponential profile extending to 3 times scale length.
The images are then convolved with a Moffat point spread function  ($\beta=3$ and FWHM=0.5 arcsec in
the deep field and 0.6 arcsec in the wide field), and buried in the simulated noise image.
The galaxies were randomly generated with various magnitudes and effective radii or scale lengths,
and analyzed with SExtractor in the same manner as for the MODS galaxies.
The results are shown in Fig.~\ref{fig-mockr90r50}. 
Seeing and noise seriously change the observed radii specifically for galaxies with a 1/4-law profile.
The observed $r_{50}$ is significantly smaller than the intrinsic values in general.
The effect is stronger for fainter 1/4-law galaxies, whereas the effect is much smaller for exponential galaxies.
The apparent smaller size than the intrinsic size would be due to the fact that the faint and large envelope 
of 1/4-law galaxies is hidden in background noise.
We should be aware of this effect if the observation is not deep enough for early-type 
galaxies (see also Hopkins et al. 2009b;  Mancini et al. 2010).
The effect will be discussed in more detail in the following section.
It is noted that Williams et al. (2010) claimed no systematic effects on $r_e$ for $m_K\la24$, using the
shallower MODS images. 

The radii, $r_{50}$ and $r_{90}$, for small galaxies are affected by smearing due to seeing effect 
and the effect depends on galaxy morphology and size.
Therefore, we obtained Eq. 1 to correct the size effect by fitting the deviation of the size as shown in 
Fig.~\ref{fig-mockr90r50} (dash line).
However, it is hard to examine the shapes of the present small or high-$z$ galaxies, so that we
do not correct the effect which depends on morphology.

\begin{eqnarray}
r_\mathrm{intrinsic} = \sqrt{r_\mathrm{observed}^2 - (\Delta r)^2}. 
\label{eq-psfcorrection}
\end{eqnarray}
We adopt $\Delta r=0.25$ and 0.35 arcsec for $r_{50}$ and 
$r_{90}$, respectively.

\begin{figure*}
\includegraphics{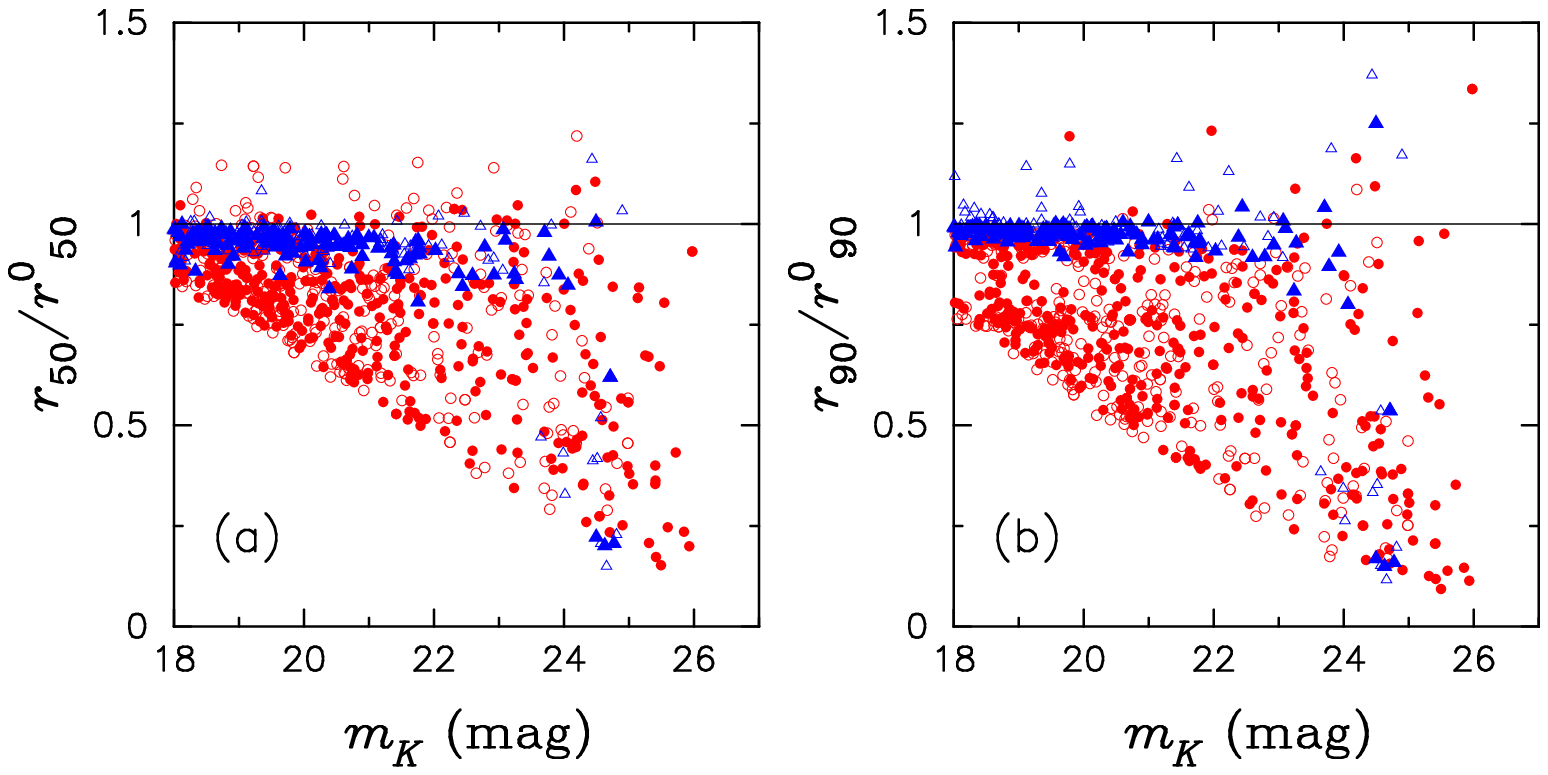}
\caption{Ratio of observed size to intrinsic for (a) $r_{50}$ and (b) $r_{90}$ as a function of the observed 
magnitude for mock galaxies.
The symbols are the same as in Fig.~\ref{fig-mockr90r50}.
}
\label{fig-comparionSize}
\end{figure*}

\begin{figure*}
\includegraphics{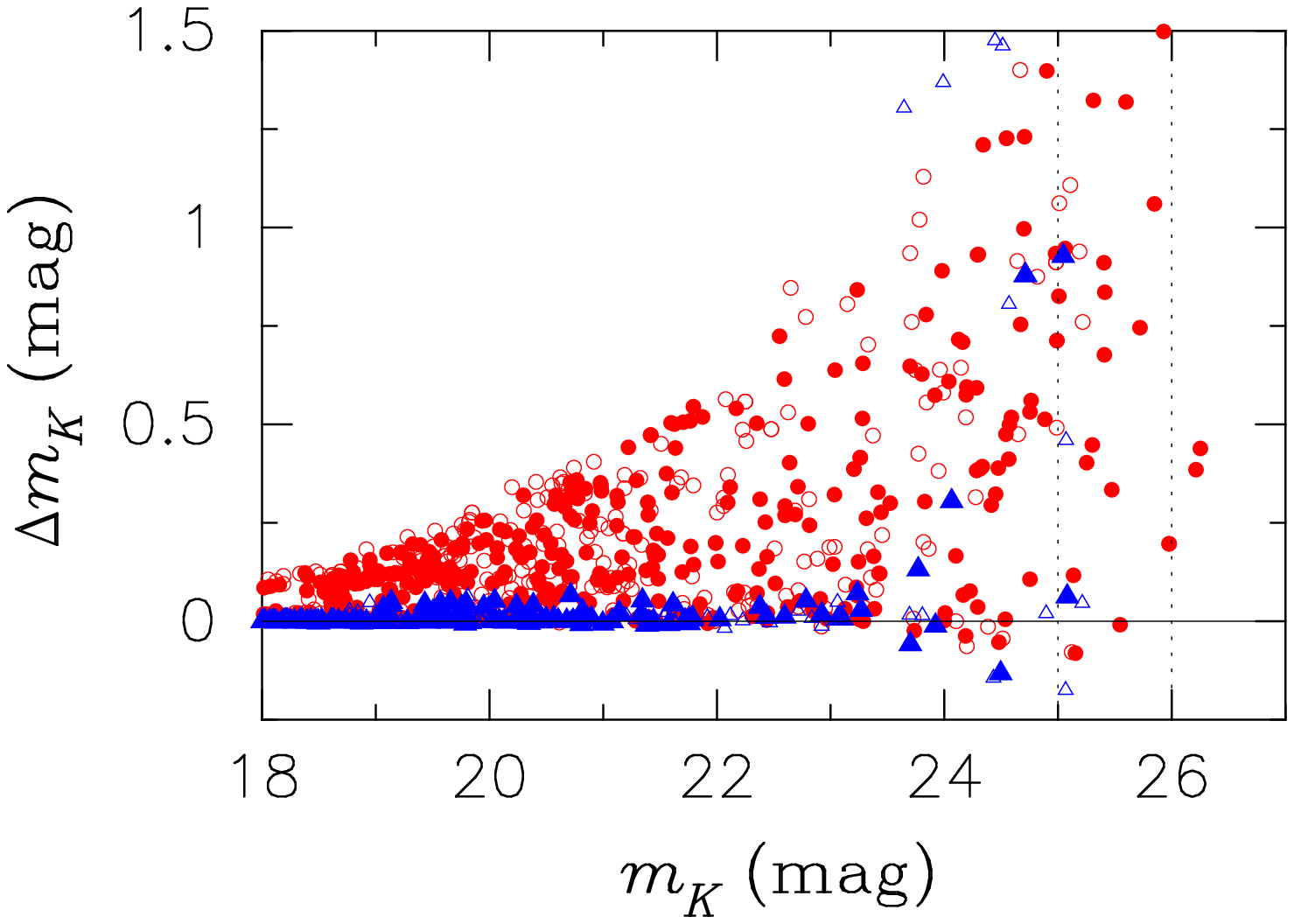}
\caption{Difference between the intrinsic and observed magnitudes (observed minus intrinsic) as a function of the observed
magnitude.
The symbols are the same as in Fig.~\ref{fig-mockr90r50}.
}
\label{fig-comparionMag}
\end{figure*}

The size errors for mock galaxies are depicted in Fig.~\ref{fig-comparionSize} as a function
of observed magnitude. 
We should take into account that much smaller values are obtained for the size of 1/4-law galaxies.
Figure \ref{fig-comparionMag} is the difference between intrinsic and observed 
magnitudes. 
At the faint limit, observed magnitudes are fainter by $1\sim1.5$ mag for 1/4-law galaxies,
while the effect is much smaller for exponential galaxies.
These results suggest that shallower observations tend to more underestimate size and flux (therefore, stellar mass).
We discuss later the influence on the results due to the systematic errors.

\subsection{Size and stellar mass relations}

\begin{figure*}
\includegraphics{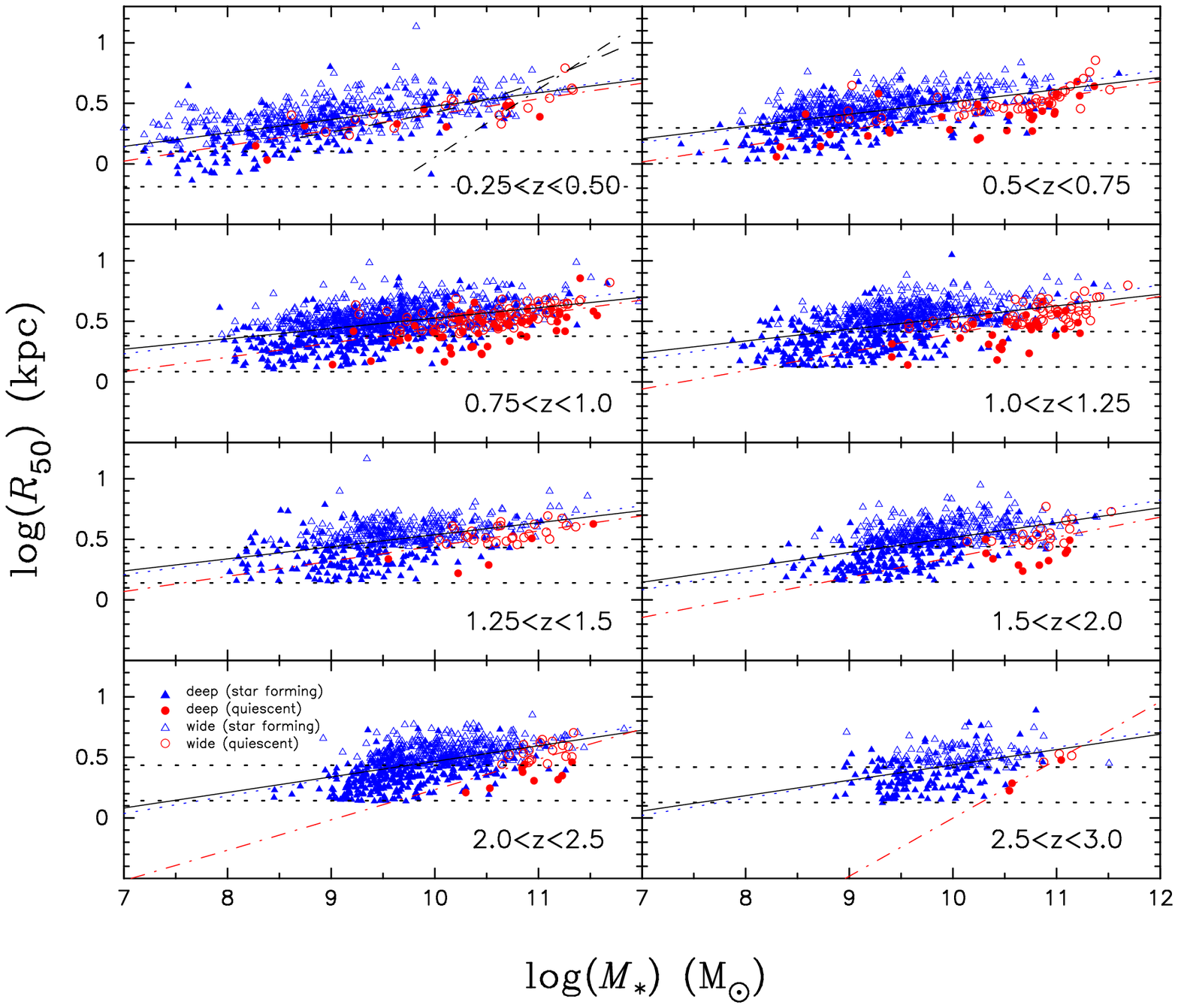}
\caption{
Physical half-light radius ($R_{50}$), corrected for PSF effects (Eq.1), as a 
function of the stellar mass ($M_*$) of the sample galaxies.
The galaxies are classified into quiescent (red) and star-forming galaxies (blue) (see text) and the fields 
(wide and deep) as noted in the bottom left frame.
Solid line is the linear regression obtained for all samples in each redshift bin.
Red dash-dot and blue dot lines are the regressions for quiescent and star-forming galaxies, respectively.
Black dashed and dash-dot lines in the $0.25<z<0.5$ bin are the results of effective radius ($r_e$) for local late- 
and early-type galaxies (Shen et al. 2003), respectively.
The horizontal dot lines are the size limits of the present observation for the wide (upper) and deep (lower) 
fields at the lower redshift boundary for each redshift bin.
}
\label{fig-massVsR50}
\end{figure*}

Using photometric (or spectroscopic, if available) redshift data, $r_{50}$ and $r_{90}$ in arcsec are 
converted to the physical size in kpc ($R_{50}$ and $R_{90}$).
The photo-$z$ accuracy of the present sample is reasonably good, $\delta z/(1+z) = -0.011$ 
($\sigma= 0.078$) (K11), so that we expect no serious influence on our result due to the photo-$z$ error.
We will confirm this later using galaxies with spectroscopic redshifts.
The results for $R_{50}$ and  $R_{90}$ are shown in Figs.~\ref{fig-massVsR50} and  \ref{fig-massVsR90} as a function of 
the stellar mass of galaxies for different redshift bins.
The unresolved galaxies are displayed in another figure (Fig.~\ref{fig-R50Reject}) for reference.
The results of effective radius ($r_e$) for local galaxies by Shen et al. (2003) are also shown in Fig.~\ref{fig-massVsR50} for comparison.
Size-mass relations for star-forming galaxies are consistent with that of late-type galaxies by Shen et al. (2003),
although our definition of half-light radius is different from that of Shen et al. (2003).
It should be noted that it would be difficult to discuss the consistency for massive early-type 
galaxies because of the small number of such galaxies in our sample.
 
Since our image quality is not high enough for classifying the galaxies 
into morphological classes at high redshifts, we divided the samples into quiescent and
star-forming galaxy groups using a two-color diagnostic plot of rest-frame $U-V$ and $V-J$ colors
(Williams et al. 2009).
The adopted selection criteria for quiescent galaxies are as follows:
\begin{eqnarray}
(U-V)>0.88(V-J)+c,
\end{eqnarray}
where $U-V$ and $V-J$ in the rest frame were obtained with the SED model fit to galaxies.
The offset, $c$, is 0.69, 0.59, and 0.49 for $0.0<z<0.5$, $0.5<z<1.0$, and $1.0<z<2.5$, respectively.
For $z>2.5$ galaxies, we applied the offset for $1<z<2.5$.
Therefore the selection would be less reliable (see Williams et al. 2009),
though the number of such galaxies is very small.
Additional criteria of $U-V>1.3$ and $V-J<1.6$ are required for quiescent galaxies at all 
redshifts to prevent contamination from unobscured and dusty star-forming galaxies, respectively.

\subsection{Stellar-mass surface density}

\begin{figure*}
\includegraphics{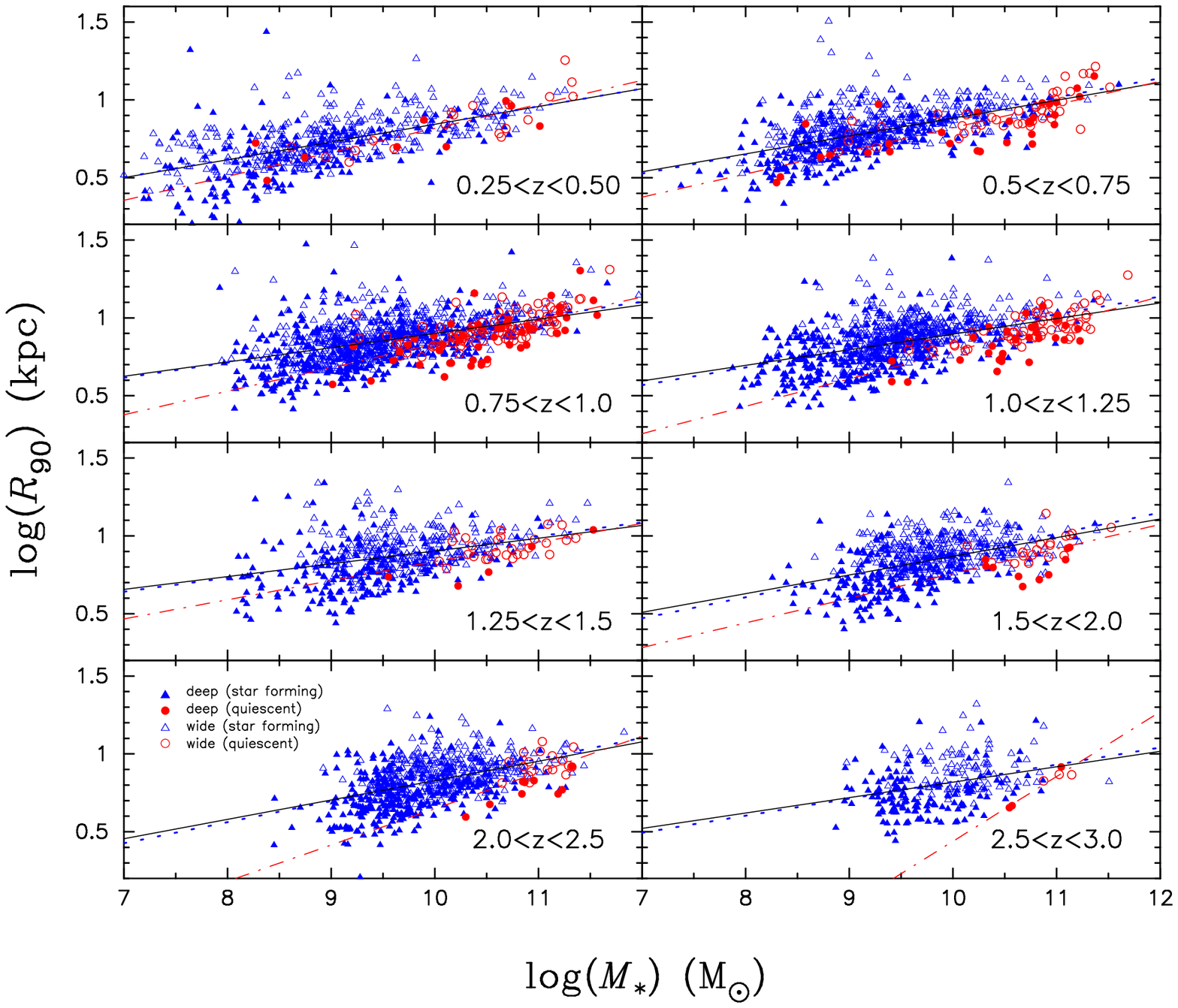}
\caption{
Same as Fig.~\ref{fig-massVsR50}, but for 90 percent-light radius ($R_{90}$).
}
\label{fig-massVsR90}
\end{figure*}

The small differences in the slope and offset of the regression lines in different redshift bins 
in Figs.~\ref{fig-massVsR50} and \ref{fig-massVsR90} would imply a universal relation between the stellar mass 
and size of galaxies, irrelevant to redshift.
In this context, we plot the stellar-mass surface density (SMSD), which is defined as $\mu_*=\mathrm{log}(M_*/\pi R^2)$, in a 
single redshift bin as a function of 
stellar mass for all galaxies at $0.25<z<3$ (Figs.~\ref{fig-SD50All} and \ref{fig-SD90All}).
The figures are essentially equivalent to the size-mass distribution.
However, we can easily identify very compact or low-surface brightness galaxies in the figures, if they exist.
For example, galaxies as compact (or diffuse) as 3 times smaller (larger) in size than the average should be located at 
$\sim$1.0 dex above (below) the average.
Such low surface brightness galaxies can be recognized at $M_*<10^{10}$ M$_\odot$ in Fig.~\ref{fig-SD90All}.

\begin{figure*}
\includegraphics{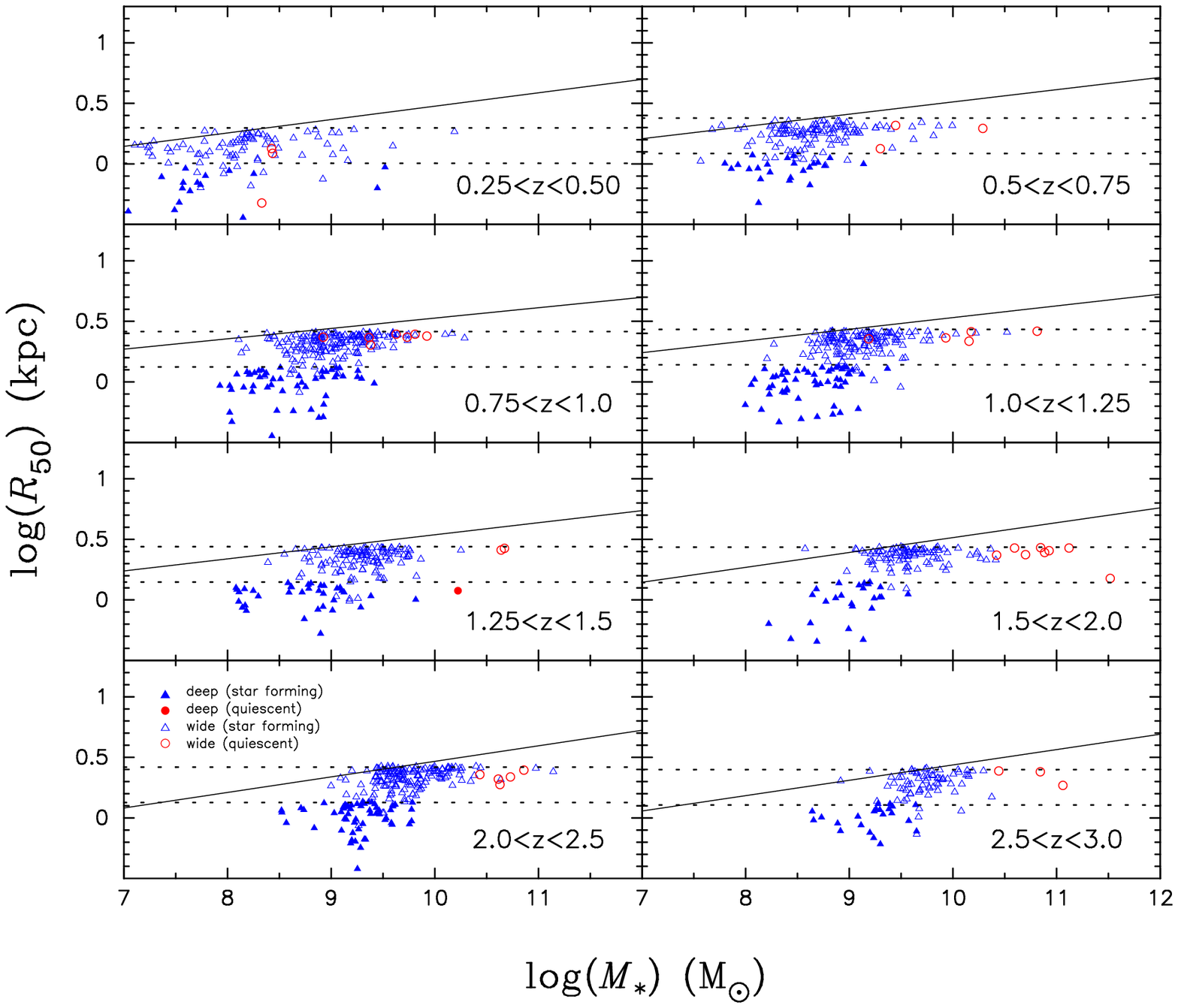}
\caption{Same as Fig.~\ref{fig-massVsR50}, but for unresolved galaxies. 
The same regression lines for all sample in each redshift bin as in Fig.~\ref{fig-massVsR50} are shown for reference.
The dotted lines are the size limits for the wide (upper) and deep (lower) 
fields at the upper redshift boundary for each redshift bin.
}
\label{fig-R50Reject}
\end{figure*}

In contrast, massive and compact galaxies are rare in the present sample.
It is noted that most of massive galaxies with $M_*>5\times 10^{10}$ M$_\odot$ should be resolved in all redshift bins
of the deep field.
We conjecture that very compact galaxies coalesce in the unresolved galaxies in Fig.~\ref{fig-R50Reject}.

\section{RESULTS}

\subsection{Linear regression of size-stellar mass relation}

Since the strong correlations of $R_{50}$ and $R_{90}$ with $M_*$ in Figs.~\ref{fig-massVsR50} and 
\ref{fig-massVsR90} are suggested, 
we obtain the least square fit between the size and mass of the galaxies with a linear regression, 
\begin{eqnarray}
\mathrm{log} R = a_r (\mathrm{log} M_* -M) + \mathrm{log} R_r^{M},
\end{eqnarray}
where $r$ is 50 for half light radius or 90 for 90\% light radius and $R_r^{M}$ is the radius at $M_*=10^{M}$ M$_\odot$.
$M$ is 10 for all and star-forming galaxies, while $M=11$ for quiescent galaxies because the quiescent galaxies 
with $M_*\la10^{10}$ M$_\odot$ were not observed at $z>1.5$ and their $a_r$ is statistically less robust due to the small numbers
of the sample in narrow mass ranges at high redshits.
We note that quiescent galaxies are located on average below the regression lines in all redshift bins, though the 
offset is smaller for $R_{90}$ than for $R_{50}$.
The fact suggests that quiescent galaxies are more compact than star-forming galaxies at a give mass.
The best-fit slopes ($a_{50}$ and $a_{90}$) and offsets ($R_{50}^{M}$ and
$R_{90}^{M}$) of the regression analysis with mean errors for all,  quiescent galaxies (QSG), and star-galaxies (SFG) are 
summarized in Table 1. 
The dispersion, $\sigma$, of the linear fit is listed in the sixth and last columns.
As the exclusion of unresolved sources could biases the results towards larger radii, 
we obtained in Table 2 the least-squares fit with the unresolved galaxies in Fig.\ref{fig-R50Reject}. 

\begin{figure*}
\includegraphics{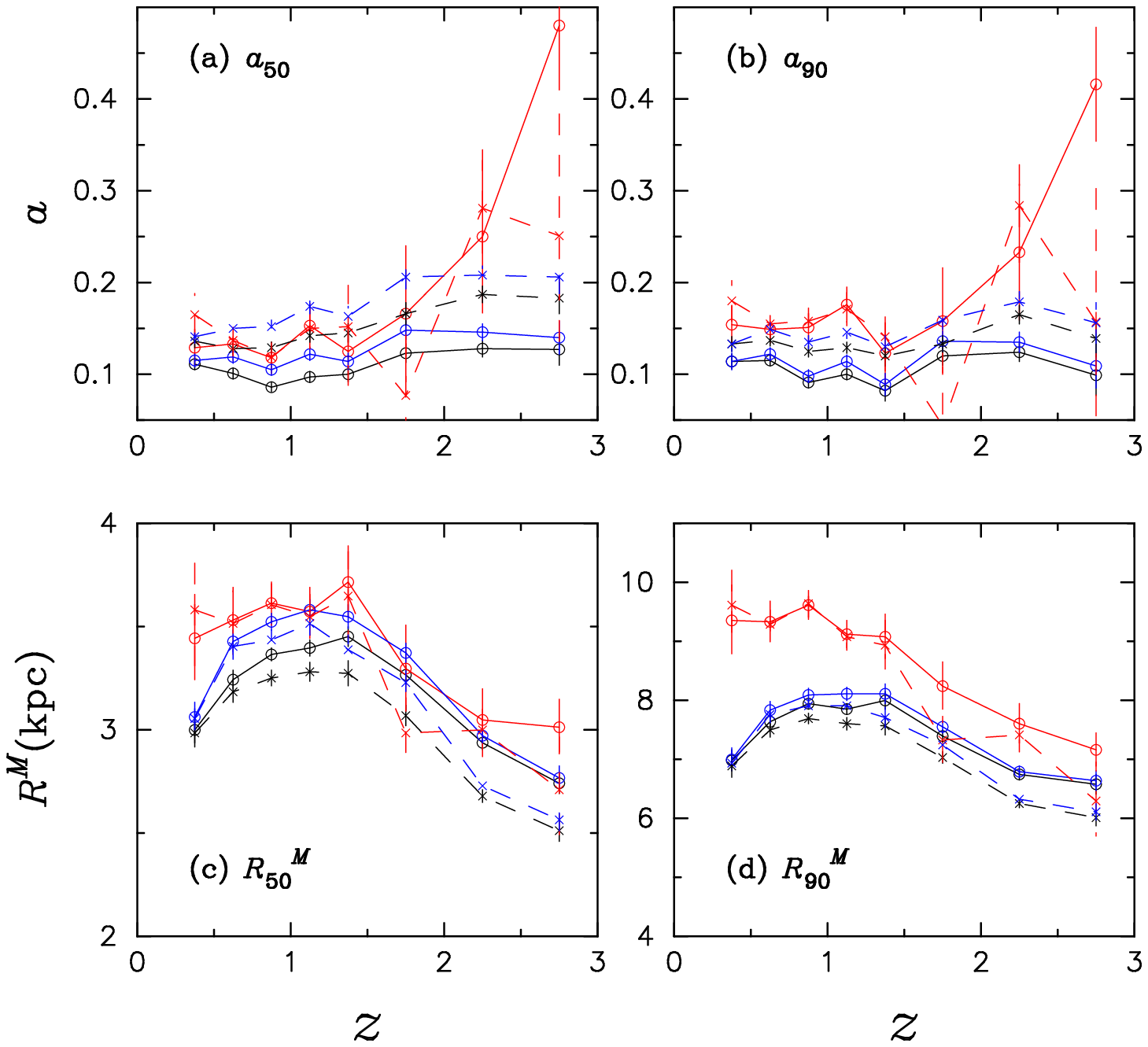}
\caption{
Evolution of the slope and offset (Tables 2 and 3) as a function of redshift.
Circles with solid line and crosses with dashed lines are the results for resolved galaxies and those with unresolved galaxies.
Black, red, and blue symbols are the samples for all, quiescent, and star-forming galaxies, respectively.
}
\label{fig-a_evolution}
\end{figure*}

We depict the evolution of the slope and offset as a function of redshift in Fig.~\ref{fig-a_evolution}.
While the slope for the quiescent galaxies are a little steeper than those for star-forming galaxies,
the figure indicates that the slopes remain within $\sim0.1-0.2$ and offsets do not significantly change from $z\sim3$ to $z\sim0.3$, 
 irrespective of the sample selection except the quiescent population at $z>2$ where the statistical
 error is very large.

\begin{figure*}
\includegraphics{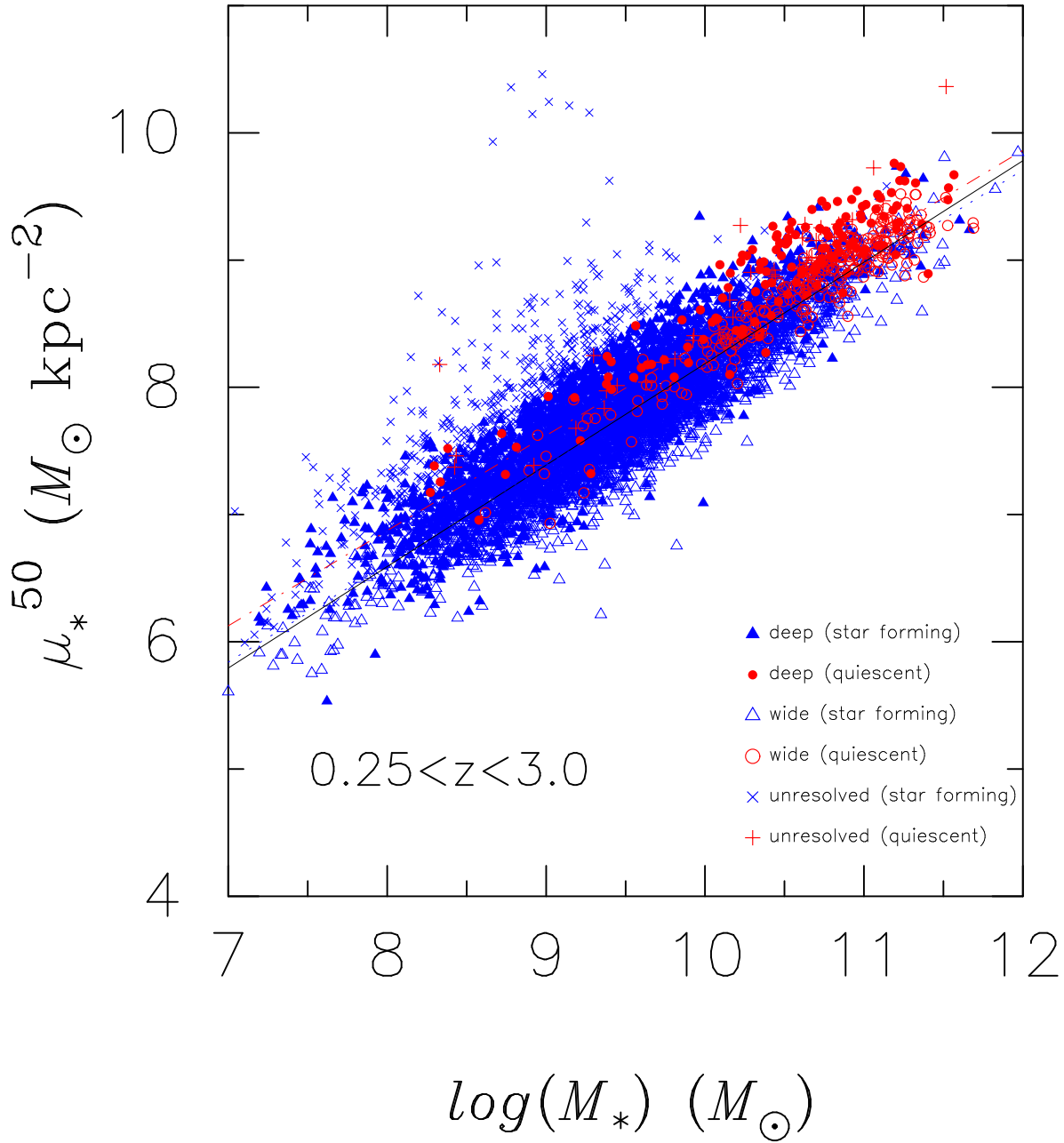}
\caption{
Correlation of stellar-mass surface density ($\mu_*$) in half-light radius ($R_{50}$) 
with stellar mass ($M_*$) of the sample galaxies in $0.3<z<3$. 
The notation of symbols are the same as in Fig.~\ref{fig-massVsR50}.
Red pluses and blue crosses show unresolved quiescent and star-forming galaxies, respectively.
Solid (black), dash-dot (red), and dot (blue) lines are the linear fits for the resolved all, 
quiescent, and star-forming galaxies.
}
\label{fig-SD50All}
\end{figure*}

\begin{figure*}
\includegraphics{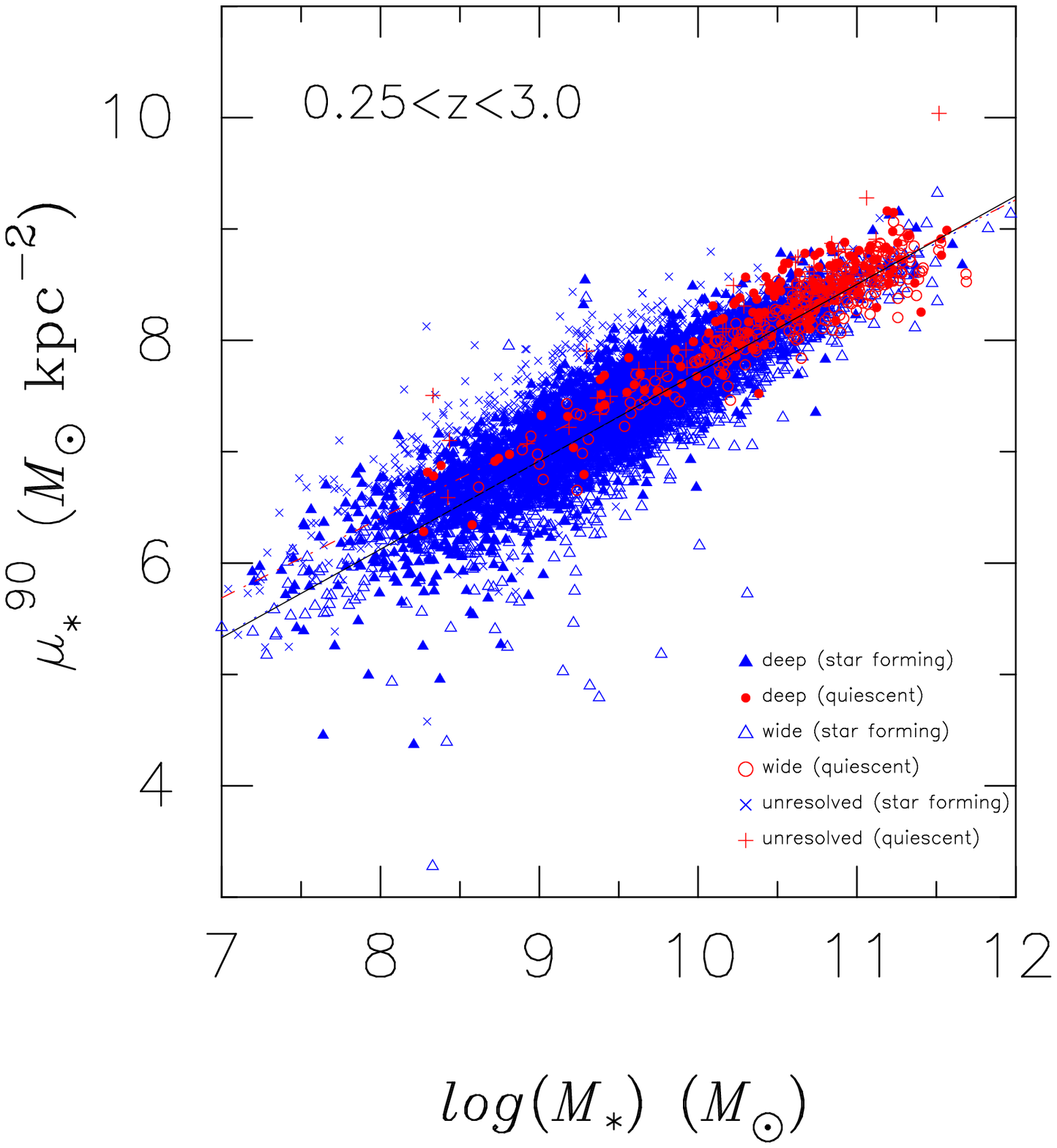}
\caption{
Same as Fig.~\ref{fig-SD50All}, but for 90 percent-light radius ($R_{90}$). 
}
\label{fig-SD90All}
\end{figure*}

For SMSD in Figs.~\ref{fig-SD50All} and \ref{fig-SD90All}, we define the regression as
\begin{eqnarray}
\mu_*=a_r^\mu \mathrm{log}(M_* -10) + \mu_r^{10},
\end{eqnarray}
where  $\mu_r^{10}$ is SMSD at $M_*=10^{10}$ M$_\odot$.
The results are shown in Table 3.

\begin{table*}
\centering
\begin{minipage}{140mm}
\caption{Linear fit of radius to stellar mass for resolved galaxies.}
  \begin{tabular}{ccrcccrccc}
  \hline
         &     &       &  $R_{50}$         &           &   &      & $R_{90}$           &         \\
\cline{3-6}\cline{7-10}
\noalign{\vspace{0.20cm}}
Redshift &  Sample &  $N$ & $a_{50}$  & $\mathrm{log} R^{M}_{50}$  & $\sigma$  & $N$ & $a_{90}$  & $\mathrm{log} R^{M}_{90}$  & $\sigma$\\
         &       & &       &   (kpc)            &           &     &      &    (kpc)           &         \\            
  \hline
$  0.25\leq z< 3.00 $ & all & 5259 & $  0.101 \pm  0.002 $ & $  0.508 \pm  0.002 $ &   0.130  &  5236 & $  0.103 \pm  0.003 $ & $  0.874 \pm  0.002 $ &   0.148 \\ 
                      & QSG &  408 & $  0.126 \pm  0.009 $ & $  0.543 \pm  0.007 $ &   0.110  &   408 & $  0.143 \pm  0.008 $ & $  0.955 \pm  0.006 $ &   0.099 \\ 
                      & SFG & 4851 & $  0.113 \pm  0.003 $ & $  0.520 \pm  0.002 $ &   0.129  &  4828 & $  0.108 \pm  0.003 $ & $  0.880 \pm  0.003 $ &   0.151 \\ 
\\                                                                                      
$  0.25\leq z< 0.50 $ & all & 543 & $  0.111 \pm  0.006 $ & $  0.477 \pm  0.009 $ &   0.133  &   542 & $  0.114 \pm  0.008 $ & $  0.844 \pm  0.011 $ &   0.168 \\ 
                      & QSG &  28 & $  0.129 \pm  0.020 $ & $  0.537 \pm  0.026 $ &   0.098  &    28 & $  0.154 \pm  0.021 $ & $  0.971 \pm  0.027 $ &   0.100 \\ 
                      & SFG & 515 & $  0.115 \pm  0.007 $ & $  0.486 \pm  0.010 $ &   0.133  &   514 & $  0.114 \pm  0.009 $ & $  0.845 \pm  0.012 $ &   0.170 \\ 
$  0.50\leq z< 0.75 $ & all & 717 & $  0.101 \pm  0.006 $ & $  0.511 \pm  0.006 $ &   0.119  &   714 & $  0.115 \pm  0.006 $ & $  0.883 \pm  0.006 $ &   0.124 \\ 
                      & QSG &  70 & $  0.133 \pm  0.018 $ & $  0.548 \pm  0.019 $ &   0.126  &    70 & $  0.149 \pm  0.015 $ & $  0.970 \pm  0.016 $ &   0.107 \\ 
                      & SFG & 647 & $  0.119 \pm  0.007 $ & $  0.535 \pm  0.007 $ &   0.113  &   644 & $  0.122 \pm  0.007 $ & $  0.894 \pm  0.008 $ &   0.124 \\ 
$  0.75\leq z< 1.00 $ & all &1083 & $  0.086 \pm  0.005 $ & $  0.527 \pm  0.004 $ &   0.123  &  1079 & $  0.091 \pm  0.006 $ & $  0.900 \pm  0.005 $ &   0.151 \\ 
                      & QSG & 134 & $  0.118 \pm  0.017 $ & $  0.558 \pm  0.012 $ &   0.104  &   134 & $  0.151 \pm  0.015 $ & $  0.983 \pm  0.011 $ &   0.095 \\ 
                      & SFG & 949 & $  0.105 \pm  0.006 $ & $  0.547 \pm  0.005 $ &   0.121  &   945 & $  0.098 \pm  0.007 $ & $  0.908 \pm  0.006 $ &   0.156 \\ 
$  1.00\leq z< 1.25 $ & all & 888 & $  0.097 \pm  0.006 $ & $  0.531 \pm  0.005 $ &   0.130  &   885 & $  0.100 \pm  0.006 $ & $  0.895 \pm  0.005 $ &   0.124 \\ 
                      & QSG &  83 & $  0.153 \pm  0.024 $ & $  0.553 \pm  0.014 $ &   0.105  &    83 & $  0.176 \pm  0.019 $ & $  0.960 \pm  0.011 $ &   0.085 \\ 
                      & SFG & 805 & $  0.122 \pm  0.007 $ & $  0.554 \pm  0.006 $ &   0.128  &   802 & $  0.114 \pm  0.007 $ & $  0.909 \pm  0.006 $ &   0.125 \\ 
$  1.25\leq z< 1.50 $ & all & 519 & $  0.100 \pm  0.008 $ & $  0.538 \pm  0.006 $ &   0.127  &   514 & $  0.082 \pm  0.011 $ & $  0.903 \pm  0.008 $ &   0.161 \\ 
                      & QSG &  32 & $  0.125 \pm  0.037 $ & $  0.570 \pm  0.020 $ &   0.089  &    32 & $  0.123 \pm  0.033 $ & $  0.958 \pm  0.018 $ &   0.079 \\ 
                      & SFG & 487 & $  0.114 \pm  0.009 $ & $  0.550 \pm  0.007 $ &   0.127  &   482 & $  0.089 \pm  0.012 $ & $  0.909 \pm  0.009 $ &   0.165 \\ 
$  1.50\leq z< 2.00 $ & all & 555 & $  0.123 \pm  0.010 $ & $  0.514 \pm  0.006 $ &   0.128  &   554 & $  0.120 \pm  0.010 $ & $  0.869 \pm  0.006 $ &   0.128 \\ 
                      & QSG &  28 & $  0.166 \pm  0.074 $ & $  0.518 \pm  0.027 $ &   0.121  &    28 & $  0.158 \pm  0.058 $ & $  0.916 \pm  0.021 $ &   0.094 \\ 
                      & SFG & 527 & $  0.148 \pm  0.010 $ & $  0.528 \pm  0.006 $ &   0.124  &   526 & $  0.136 \pm  0.011 $ & $  0.878 \pm  0.006 $ &   0.128 \\ 
$  2.00\leq z< 2.50 $ & all & 700 & $  0.128 \pm  0.009 $ & $  0.468 \pm  0.005 $ &   0.121  &   696 & $  0.124 \pm  0.010 $ & $  0.829 \pm  0.006 $ &   0.146 \\ 
                      & QSG &  27 & $  0.250 \pm  0.083 $ & $  0.484 \pm  0.021 $ &   0.109  &    27 & $  0.233 \pm  0.074 $ & $  0.881 \pm  0.019 $ &   0.097 \\ 
                      & SFG & 673 & $  0.146 \pm  0.009 $ & $  0.473 \pm  0.005 $ &   0.119  &   669 & $  0.135 \pm  0.011 $ & $  0.832 \pm  0.006 $ &   0.147 \\ 
$  2.50\leq z< 3.00 $ & all & 254 & $  0.127 \pm  0.017 $ & $  0.438 \pm  0.008 $ &   0.130  &   252 & $  0.099 \pm  0.022 $ & $  0.818 \pm  0.011 $ &   0.166 \\ 
                      & QSG &   6 & $  0.480 \pm  0.070 $ & $  0.479 \pm  0.019 $ &   0.040  &     6 & $  0.416 \pm  0.062 $ & $  0.855 \pm  0.017 $ &   0.035 \\ 
                      & SFG & 248 & $  0.140 \pm  0.018 $ & $  0.442 \pm  0.009 $ &   0.129  &   246 & $  0.109 \pm  0.024 $ & $  0.822 \pm  0.011 $ &   0.167 \\                                                                                                                                      
\hline
\end{tabular}
\end{minipage}
\end{table*}

\begin{table*}
\centering
\begin{minipage}{140mm}
\caption{Linear fit of radius to stellar mass with unresolved galaxies}
  \begin{tabular}{ccrcccrccc}
  \hline
         &     &       &  $R_{50}$         &           &   &      & $R_{90}$           &         \\
\cline{3-6}\cline{7-10}
\noalign{\vspace{0.20cm}}
Redshift &  Sample &  $N$ & $a_{50}$  & $\mathrm{log} R^{M}_{50}$  & $\sigma$  & $N$ & $a_{90}$  & $\mathrm{log} R^{M}_{90}$  & $\sigma$\\
         &     &   &      &   (kpc)            &           &     &      &    (kpc)           &         \\            
\hline
$  0.25\leq z< 3.00 $ & all & 6532 & $  0.131 \pm  0.003 $ & $  0.484 \pm  0.002 $ &   0.161  &  6502 & $  0.125 \pm  0.003 $ & $  0.852 \pm  0.002 $ &   0.161 \\ 
                      & QSG &  445 & $  0.132 \pm  0.008 $ & $  0.536 \pm  0.007 $ &   0.116  &   445 & $  0.147 \pm  0.008 $ & $  0.946 \pm  0.006 $ &   0.108 \\ 
                      & SFG & 6087 & $  0.144 \pm  0.003 $ & $  0.497 \pm  0.003 $ &   0.162  &  6057 & $  0.130 \pm  0.003 $ & $  0.859 \pm  0.003 $ &   0.164 \\ 
\\                                                                                           
$  0.25\leq z< 0.50 $ & all &  646 & $  0.136 \pm  0.007 $ & $  0.475 \pm  0.010 $ &   0.156  &   645 & $  0.133 \pm  0.008 $ & $  0.838 \pm  0.012 $ &   0.181 \\ 
                      & QSG &   31 & $  0.165 \pm  0.023 $ & $  0.554 \pm  0.034 $ &   0.129  &    31 & $  0.180 \pm  0.022 $ & $  0.983 \pm  0.033 $ &   0.124 \\ 
                      & SFG &  615 & $  0.141 \pm  0.007 $ & $  0.485 \pm  0.011 $ &   0.156  &   614 & $  0.133 \pm  0.008 $ & $  0.841 \pm  0.013 $ &   0.183 \\ 
$  0.50\leq z< 0.75 $ & all &  858 & $  0.128 \pm  0.006 $ & $  0.503 \pm  0.007 $ &   0.136  &   855 & $  0.137 \pm  0.006 $ & $  0.875 \pm  0.007 $ &   0.134 \\ 
                      & QSG &   73 & $  0.137 \pm  0.017 $ & $  0.546 \pm  0.019 $ &   0.127  &    73 & $  0.155 \pm  0.015 $ & $  0.968 \pm  0.016 $ &   0.111 \\ 
                      & SFG &  785 & $  0.150 \pm  0.007 $ & $  0.532 \pm  0.008 $ &   0.133  &   782 & $  0.149 \pm  0.007 $ & $  0.890 \pm  0.008 $ &   0.135 \\ 
$  0.75\leq z< 1.00 $ & all & 1313 & $  0.129 \pm  0.006 $ & $  0.512 \pm  0.005 $ &   0.155  &  1308 & $  0.125 \pm  0.006 $ & $  0.886 \pm  0.005 $ &   0.161 \\ 
                      & QSG &  141 & $  0.119 \pm  0.015 $ & $  0.557 \pm  0.012 $ &   0.102  &   141 & $  0.158 \pm  0.014 $ & $  0.984 \pm  0.011 $ &   0.094 \\ 
                      & SFG & 1172 & $  0.152 \pm  0.007 $ & $  0.536 \pm  0.006 $ &   0.157  &  1167 & $  0.135 \pm  0.007 $ & $  0.898 \pm  0.007 $ &   0.166 \\ 
$  1.00\leq z< 1.25 $ & all & 1082 & $  0.142 \pm  0.007 $ & $  0.516 \pm  0.006 $ &   0.164  &  1077 & $  0.129 \pm  0.006 $ & $  0.881 \pm  0.006 $ &   0.146 \\ 
                      & QSG &   88 & $  0.150 \pm  0.022 $ & $  0.550 \pm  0.013 $ &   0.103  &    88 & $  0.171 \pm  0.018 $ & $  0.958 \pm  0.011 $ &   0.084 \\ 
                      & SFG &  994 & $  0.174 \pm  0.008 $ & $  0.546 \pm  0.007 $ &   0.164  &   989 & $  0.146 \pm  0.007 $ & $  0.898 \pm  0.007 $ &   0.148 \\ 
$  1.25\leq z< 1.50 $ & all &  675 & $  0.145 \pm  0.010 $ & $  0.515 \pm  0.008 $ &   0.159  &   668 & $  0.120 \pm  0.011 $ & $  0.879 \pm  0.009 $ &   0.176 \\ 
                      & QSG &   35 & $  0.152 \pm  0.045 $ & $  0.562 \pm  0.025 $ &   0.111  &    35 & $  0.141 \pm  0.037 $ & $  0.951 \pm  0.020 $ &   0.091 \\ 
                      & SFG &  640 & $  0.163 \pm  0.011 $ & $  0.530 \pm  0.009 $ &   0.160  &   633 & $  0.130 \pm  0.012 $ & $  0.887 \pm  0.010 $ &   0.179 \\ 
$  1.50\leq z< 2.00 $ & all &  685 & $  0.166 \pm  0.012 $ & $  0.487 \pm  0.007 $ &   0.174  &   684 & $  0.133 \pm  0.009 $ & $  0.847 \pm  0.006 $ &   0.139 \\ 
                      & QSG &   36 & $  0.077 \pm  0.069 $ & $  0.475 \pm  0.024 $ &   0.128  &    36 & $  0.044 \pm  0.064 $ & $  0.865 \pm  0.023 $ &   0.120 \\ 
                      & SFG &  649 & $  0.206 \pm  0.013 $ & $  0.509 \pm  0.008 $ &   0.171  &   648 & $  0.159 \pm  0.010 $ & $  0.860 \pm  0.006 $ &   0.138 \\ 
$  2.00\leq z< 2.50 $ & all &  923 & $  0.187 \pm  0.009 $ & $  0.428 \pm  0.005 $ &   0.152  &   918 & $  0.165 \pm  0.010 $ & $  0.796 \pm  0.005 $ &   0.156 \\ 
                      & QSG &   32 & $  0.281 \pm  0.068 $ & $  0.477 \pm  0.019 $ &   0.103  &    32 & $  0.284 \pm  0.062 $ & $  0.870 \pm  0.017 $ &   0.094 \\ 
                      & SFG &  891 & $  0.208 \pm  0.010 $ & $  0.436 \pm  0.005 $ &   0.151  &   886 & $  0.179 \pm  0.011 $ & $  0.801 \pm  0.005 $ &   0.157 \\ 
$  2.50\leq z< 3.00 $ & all &  350 & $  0.183 \pm  0.017 $ & $  0.400 \pm  0.009 $ &   0.150  &   347 & $  0.139 \pm  0.020 $ & $  0.779 \pm  0.010 $ &   0.175 \\ 
                      & QSG &    9 & $  0.251 \pm  0.133 $ & $  0.433 \pm  0.039 $ &   0.097  &     9 & $  0.157 \pm  0.148 $ & $  0.799 \pm  0.043 $ &   0.109 \\ 
                      & SFG &  341 & $  0.206 \pm  0.018 $ & $  0.409 \pm  0.009 $ &   0.149  &   338 & $  0.156 \pm  0.022 $ & $  0.786 \pm  0.010 $ &   0.176 \\ 
\hline
\end{tabular}                                                                           
\end{minipage}
\end{table*}

\begin{table*}
\centering
\begin{minipage}{140mm}
\caption{Linear fit of surface mass density of galaxies in $0.25<z<3.0$.}
  \begin{tabular}{ccrccrccccc}
  \hline
         &     &       &  $R_{50}$         &           &   &      & $R_{90}$           &         \\
\cline{3-6}\cline{7-10}
\noalign{\vspace{0.20cm}}
Selection$^{a}$ & Sample &   N & $a_{50}^{\mu} $  & $\mu_{50}^{10}$                & $\sigma$  & N & $a_{90}^{\mu}$  & $\mu_{90}^{10}$                & $\sigma$\\
            &   &   &      & ($M_\odot$ kpc$^{-2}$) &  &   &      & ($M_\odot$ kpc$^{-2}$) & \\
  \hline

 $m_K\leq 26 (25)$    &  all & 5259 & $  0.80 \pm  0.00 $ & $  8.19 \pm  0.00 $ &   0.26 &   5236 & $  0.79 \pm  0.01 $ & $  7.71 \pm  0.00 $ &   0.30 \\ 
                      &  QSG &  408 & $  0.75 \pm  0.02 $ & $  8.37 \pm  0.01 $ &   0.22 &    408 & $  0.71 \pm  0.02 $ & $  7.83 \pm  0.01 $ &   0.20 \\ 
                      &  SFG & 4851 & $  0.77 \pm  0.01 $ & $  8.16 \pm  0.00 $ &   0.26 &   4828 & $  0.78 \pm  0.01 $ & $  7.70 \pm  0.01 $ &   0.30 \\ 
 with unresolved galaxies &  all & 6532 & $  0.74 \pm  0.01 $ & $  8.23 \pm  0.00 $ &   0.32 &  6502 & $  0.75 \pm  0.01 $ & $  7.75 \pm  0.00 $ &   0.32 \\ 
                          &  QSG &  445 & $  0.74 \pm  0.02 $ & $  8.39 \pm  0.01 $ &   0.23 &   445 & $  0.71 \pm  0.02 $ & $  7.86 \pm  0.01 $ &   0.22 \\ 
                          &  SFG & 6087 & $  0.71 \pm  0.01 $ & $  8.21 \pm  0.01 $ &   0.32 &  6057 & $  0.74 \pm  0.01 $ & $  7.74 \pm  0.01 $ &   0.33 \\ 
 $m_K\leq 25 (24)$  &   &  3980 &   $  0.80 \pm 0.01  $ &   $  8.19  \pm 0.00 $  &   0.25 &  3972  & $  0.80 \pm 0.01 $  & $  7.71 \pm  0.00 $  &  0.28 \\
 $m_K\leq 24 (23)$  &   & 2378 &   $  0.82 \pm 0.01  $ &   $  8.20  \pm 0.01 $  &   0.25 &  2378  & $  0.79 \pm 0.01 $  & $  7.71 \pm  0.00 $  &  0.23 \\
 spectroscopic redshift  &   &  1610 &   $  0.81 \pm 0.01  $ &   $  8.16  \pm 0.01 $  &   0.24 &   1610  & $  0.77 \pm 0.01 $  & $  7.69 \pm  0.01 $  &  0.22 \\
                                                                                                                                           
\hline
\noalign {\vspace{0.20cm}}
\multicolumn{8}{l}{\small $^{a)}$ Selection for the deep field (parentheses are the sample for the wide field)} \
\end{tabular}
\end{minipage}
\end{table*}

\subsection{Error estimate}
The errors in stellar mass mainly originated from 
SED fitting and photometric errors of the observations.
In the course of $\chi^2$ fitting to obtain the best SED model with various parameters 
(e.g., star-formation time scale, photometric redshift for galaxies with no spectroscopic redshift available, 
age, extinction, and metallicity), the probability distributions of stellar mass can be 
calculated, where the photometry and photometric-redshift errors are included.
The error for the observed SMSD is dominated by the error for the stellar mass.
See more details for the error estimate in K09 and Ichikawa et al. (2010)

The size and magnitude errors in accordance with galaxy magnitude have been examined using mock galaxies 
(Figs.~\ref{fig-comparionSize} and \ref{fig-comparionMag}).
If all galaxies have a shape of 1/4-law, the error would be significant.
Recalling the size error due to the morphology strongly depends on magnitude, we examine the effect 
by confining the sample to bright galaxies, where the systematic error becomes smaller.
We selected the samples with the magnitude limits, $m_K=24$ (23), 25(24), 26(25) for the deep (wide) fields,
then obtained the regression lines again.
The results are compared in Table 3.
It should be noted that the section does not change the result.

One would be concerned about the reliance on photometric redshift estimates for high redshift galaxies
 (e.g., Mosleh et al. 2011).
In order to examine the error, we selected only the galaxies with spectroscopic 
redshift available and compared the result with the photo-$z$ sample.
Although the spectroscopic samples are limited to lower redshift, the result does not change our conclusion
(Table 3).

\begin{figure*}
\includegraphics{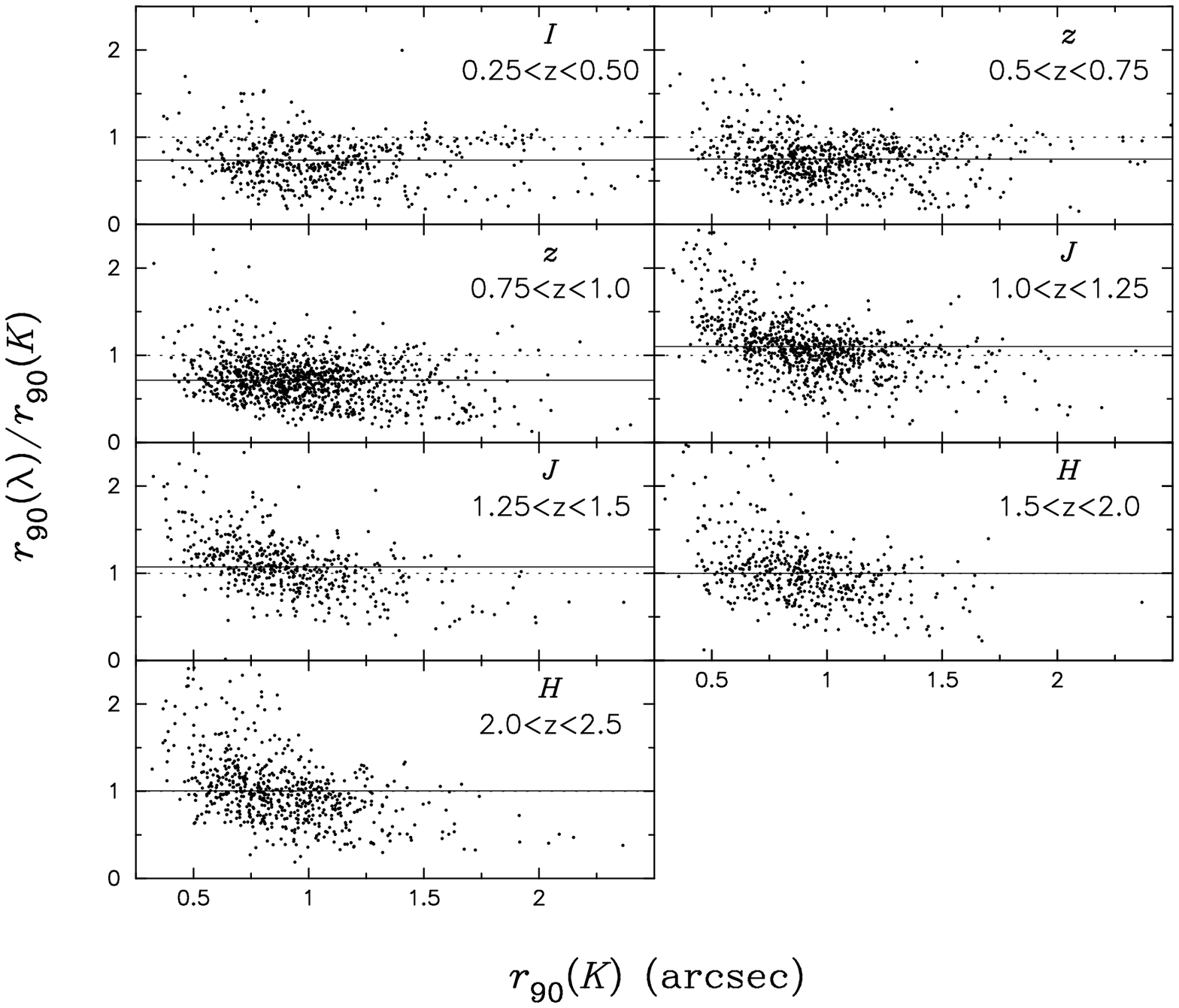}
\caption{Ratio of 90 percent-light radius ($r^\lambda_{90}$) at the nearest rest-frame $V$ band to 
that of $K$ band  ($r^K_{90}$).
Solid line represents the average for the samples in each redshift bin and dotted line is the equal 
value.
The nearest band is depicted at the upper corner for each redshift bin.
For $I$ and $z$ bands, we use the HST images binned to  0.117 arcsec pixel$^{-1}$.
}
\label{fig-R90vsACS}
\end{figure*}

Due to possible color gradients in galaxies, it would be best to use images of the same rest-frame band 
with comparable $S/N$ for measuring the size of galaxies at all redshifts.
Nevertheless, we used the radius of galaxies measured on $K$-band image, because 
it is deepest among the images we used and because it is the rest-frame optical or longer
 band at $z\la3$.
If the central region of galaxies are younger (bluer) than the outer region, the mass-weighted radius could be larger than the 
luminosity-weighted radius.
If it is older (redder), the result would be vice versa. 
Star forming galaxies sometimes show strong morphological variation between observed wavelengths.
Bond, Gawiser \& Koekemoer (2011) reported that this was  not generically accompanied by a large difference in half-light radius.
Barden et al. (2005) measured the disk scale lengths of local galaxies in various bands.
The average size is about 10 percent larger in $V$ band than in $K$ band.
On the other hand, MacArthur \& Courteau (2003) showed the contrary result that the distribution of disc scale lengths 
was a decreasing trend with increasing wavelengths (see also Cassata et al. 2010 for early-type galaxies at $z\sim2$).

To investigate the effect of the color gradient, we compare the 90 present-light radius of galaxies measured on the nearest rest-frame $V$ 
band for each redshift bin.
The sizes of the present samples were obtained on ACS $I$, $z$, and MOIRCS $J$, $H$ bands and compared with that of $K$ band in Fig.~\ref{fig-R90vsACS}.
We used ACS images binned to 0.117 arcsec per pixel, keeping the original image resolution.
Fig.~\ref{fig-R90vsACS} demonstrates that the sizes of ACS in $I$ and $z$ bands are  systematically $\sim25$ percent smaller 
than that in $K$ band. 

The convolution of the ACS $I$-band image with PSF and seeing enhances low surface brightness details.
If the ACS images are convolved with a gaussian (FWHM=2 pixel, 0.234 arcsec), the difference is decreased 
to  $\sim$15 percent.
As the convolution enhances the galaxy edge of low surface brightness, it tends to give a larger galaxy size 
(see also fig. 7 of Mancini et al. 2010).
It should be noted that the $J$ and $H$ images gives larger image size for smaller galaxies.
The size strongly depends on the depth and the seeing size.

\subsection{Size evolution}

\begin{figure*}
\includegraphics{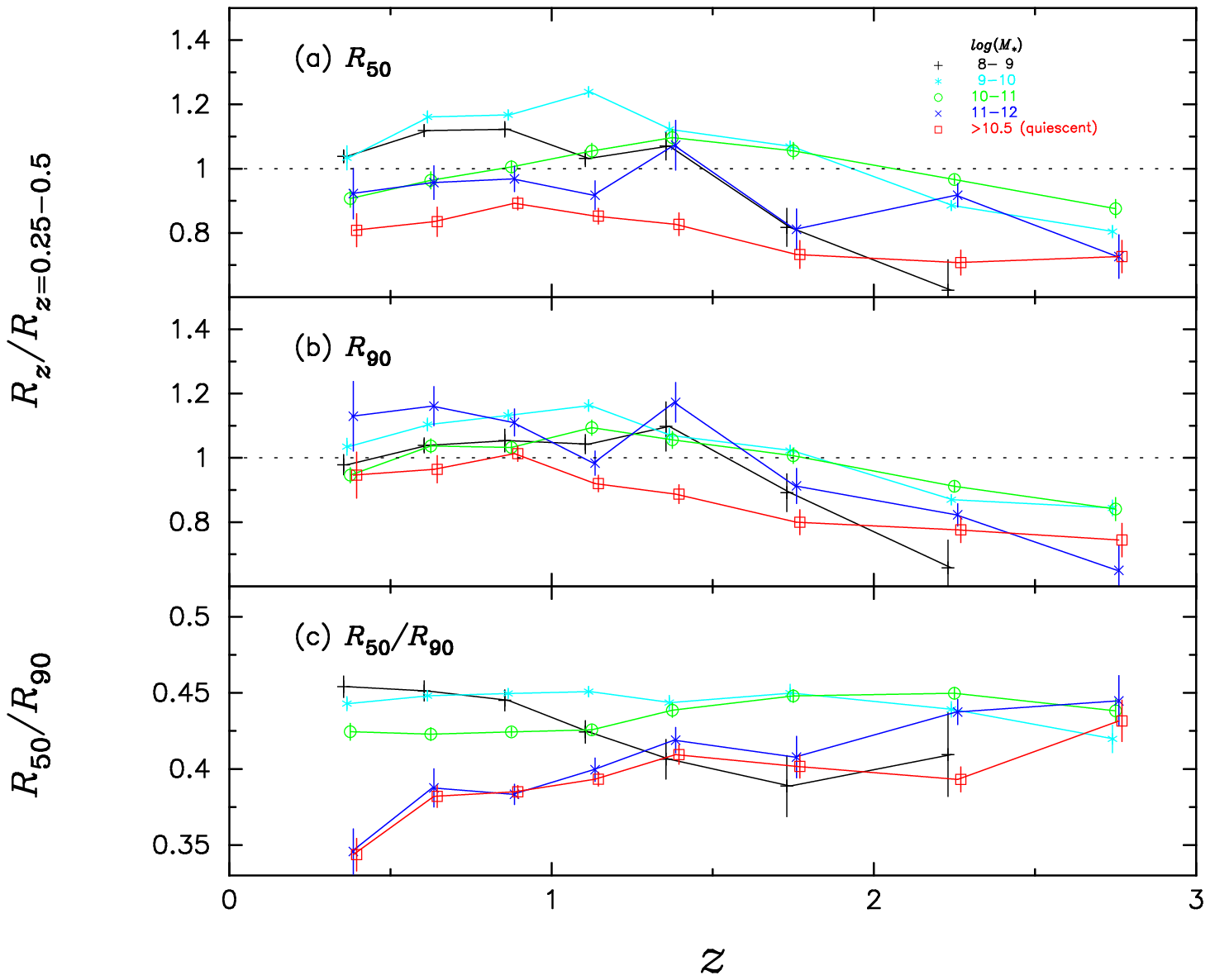}
\caption{Evolution of galaxy sizes as a function of redshift in stellar mass bins.
The ordinate is the difference of the median sizes, (a) $R_{50}$ and (b) $R_{90}$,  from the regression line 
obtained for the galaxies in $0.25<z<0.5$ (Table 2). (c) Median size ratio $R_{50}/R_{90}$.
Unresolved galaxies are included.
The error bar is the mean error of the average.
}
\label{fig-R_evolution}
\end{figure*}

Although we have found the universal relation between $M_*$ and SMSD (or size), which does not depend strongly
on redshift, we have good reason to expect offsets from the relation for some galaxy populations (e.g., 
massive compact galaxies) in certain mass and redshift ranges.
The regression analysis could be more weighted on the numerous less massive populations.
Therefore, it is likely that there are small (but not highly significant) offsets between galaxy populations.
The small offsets would account for the size evolution of such populations.
Using the  $M_*$--$R$ relation obtained for the galaxies at $0.25<z<0.5$ (Tables 1 and 2) as a reference, we examined
the deviation of the median size of galaxies in the mass and redshift bins from the reference.
The evolution of the massive ($M_*>10^{10.5}$M$_\odot$) quiescent galaxies, defined in a $U-V$ and $V-J$ diagram (Williams et al. 2010),
is also obtained.
(We use median values to avoid unreasonable contributions from outliers with large deviation.)
We show the results in Fig.~\ref{fig-R_evolution}, where unresolved galaxies are included.
In addition, the median values of $R_{50}/R_{90}$, which represent a sort of {\it compactness} of galaxies, 
are depicted in the figure to see the evolution of the compactness as a function of redshift.
We note that the result with resolved galaxies are in good agreement with that with unresolved galaxies.

\section{DISCUSSION}

Using the deepest $K$-band image, 
we obtained scaling relations between stellar-mass ($M_*$) and size (or equivalently SMSD) as a function of redshift
for galaxies at $0.3<z<3$ in a wide mass range of $M_*\sim10^{8}-10^{11}$ M$_\odot$.
We defined the radii encircling half- ($R_{50}$) and 90 percent-light ($R_{90}$) in a circular aperture.
The depth was found to be crucial for proper measure of the size and mass for less massive galaxies dimmed due to cosmic expansion.
We conclude that there is  no strong evidence for the size evolution at a given mass over the redshift range, irrespective of galaxy
populations (star-forming, quiescent).
The size-mass relation ($R\propto M_*^\alpha$) have a universal slope ($\alpha \sim 0.1-0.2$) and small offsets ($\la 50$ percent). 
In other words, as galaxies grow in mass through star formation or merging process,
their sizes evolve in such a way that galaxies in general move along the scaling relation.
The universal relation demonstrates that the stellar mass in galaxies with the same stellar mass was 
built up on average in a similar manner over cosmic times.
The trend is insensitive to the stellar populations.
Our result is comparable with  the $r_e\propto M_*^{0.15}$ found by Shen et al. (2003) for low-mass galaxies 
($M_*<10^{10.6}$ M$_\odot$).
On the other hand, it is in contrast to the steeper relation for massive galaxies
obtained by previous studies (e.g., Shen 2003; Bernardi et al. 2011).
The scaling relations are much tighter ($\sigma\sim0.12$--0.17 dex) than those ($\sigma\sim0.3$--0.5 dex) 
of previous studies in the local universe (e.g., Shen et al. 2003) or at high redshifts (e.g., Franx et al. 2008).

The weak growth of the galaxy size, irrelevant to galaxy mass, is in disagreement with the scenario that more massive 
galaxies rapidly changed their sizes.
If massive high-$z$ galaxies are several times smaller than local galaxies with comparable mass, they  
should be located well above the stellar mass vs. SMSD relations.
Several candidates of such compact galaxies with $M_*>10^{11}$ possibly coalesce as unresolved galaxies  
in Figs.~\ref{fig-SD50All} and \ref{fig-SD90All}.
However, such galaxies are found to be few.
Our finding is in contrast with the previous results for quiescent early-type massive galaxies (e.g., Zirm et al. 2007; Williams et al. 2010).
With regard to SMSD, the average SMSDs in $R_{50}$ for our sample galaxies with  $M_*>10^{10}$ M$_\odot$ 
are found not to evolve with redshift (log $\mu_{R_{50}}=8.53$ -- 8.70). 
The average SMSD for all galaxies at $0.25<z<3.0$ is $8.60 \pm 0.01$, which is very consistent
with the result, $0.85 \pm 0.03$, for disk galaxies with $M_*>10^{10}$ obtained by Barden et al (2005),
though our sample could be a mixture of disk and elliptical galaxies.
However, due to small volume of the present study, the results could be subject to statistical 
uncertainties.

The weak size evolution presented in the present study is reconciled with the surface brightness (SB) evolution.
Ichikawa et al. (2010) presented a universal linear correlation between the stellar mass and SB 
in rest-frame $V$ and $z$ bands for galaxies at $0.3<z<3$, using the same sample as the present study.
The correlation has a nearly constant slope, independent of redshift and color of galaxies in the 
rest-$z$ frame.
In contrast, SB shows a strong dependence on redshift for a given stellar mass.
It evolves as $(1+z)^{-2.0\sim-0.8}$.
The increase in the luminosity with redshift is estimated by using the expected luminosity 
evolution from a single burst at $z=4$ for massive galaxies ($M_*>10^{10}$)
and constant star formation for blue samples ($U-V<0$) for less massive galaxies.
The redshift dependence of SB evolution is well explained by the pure luminosity evolution of galaxies out to $z\sim3$
without the need of the size evolution, which supports the results for young early-type galaxies  (e.g., 
Saracco et al. 2011) and for disk galaxies at $z\la1$ (Barden et al. 2005) .

Nevertheless, there are some populations having a small offset from the universal scaling relation.
We see more or less evolution in the size and compactness for such populations as a function of redshift in 
Fig.~\ref{fig-R_evolution}. 
The figure suggests a moderate increase of 30--50 percent for $R_{50}$ and $R_{90}$ for less massive galaxies
 ($M_*<10^{10}$ M$_\odot$) from $z\sim3$ to $z\sim1$, while 
the sizes remains unchanged or slightly decrease towards $z\sim0.3$.
For massive galaxies ($M_*\ga10^{11}$ M$_\odot$), the evolution is $\sim70-80$ percent in $R_{90}$ 
from $z\sim3$ to $z\sim0.3$, though that in $R_{50}$ is weaker.
$R_{90}$  evolved as $\propto (1 + z)^{-0.5}$, 
It is noted that Trujillo et al. (2006), using the ground-based $K$-band data comparable with our depth, 
but in a smaller region (6.3 arcmin$^2$), concluded that there was no evidence for significant evolution 
for the size-mass relation and that there was small increase (29 percent) since $z\sim2.5$ for 
the most massive bin $M_*>4\times10^{10}$ M$_\odot$.
In our sample, the average $R_{50}$ for the most massive bin of $M_*=10^{11}$--$10^{12}$ M$_\odot$ is 
larger by $\sim20-30$ percent at $z\sim0.3$ than at $z\sim2.5$,
which is in good agreement with the result of Trujillo et al. (2006), whereas it is 
much weaker than those of other studies for galaxies of similar mass 
(e.g., Franx et al. 2008; Buitrago et al. 2008; van der Wel et al. 2009;
Williams et al. 2010).
Using new deep WFC3 data, Cassata et al. (2011) and Law et al. (2012) have given 
new evidences of compact galaxies at high-z. Although the data in $H$ band is shallower by 
1 mag arcsec$^{-2}$ than our $K$-band data, their conclusion would be robust.
It would be possible that the compact galaxies are mingled in unresolved galaxies of 
the present sample or that the difference of the definition for half-right radius ($r_\mathrm{e}$ and $r_{50}$) 
gives the contradict result.

The ratio, $R_{50}/R_{90}$, in Fig.~\ref{fig-R_evolution} would give a clue to understanding
the evolution of the compactness of galaxies.
While middle and low mass galaxies shows no evolution in the compactness, the galaxies in the most massive bin
are more compact at higher redshift.
Recalling weaker evolution in $R_{50}$ than in $R_{90}$,
we infer that the compactness evolution is ascribed mainly to the expansion of the outer rim of massive galaxies.
In other words, at a given mass, massive galaxies in the local universe are more influenced  by mergers or star
formation at the outer envelopes than those at high redshifts.
It would be worth noting that the galaxies in the massive bin of the deep field are complete 
in sampling over the present redshift range and their $R_{50}$ and $R_{90}$ are well resolved in the present PSF.
However, the present observation samples a comparatively small volume and therefore, 
the results could be subject to statistical uncertainties for massive galaxies of low number density due to 
field variances.

The size evolution is often used to advocate the merging processes in the hierarchical paradigm of
galaxy formation and evolution.
Minor dry merger is a plausible mechanism for weak size evolution (e.g., Guo \& White 2008; Bezanson et al. 2009;
Naab, Johansson, \& Ostriker 2009; Hopkins et al. 2010).
In that the size evolution is stronger in massive galaxies, 
our findings are qualitatively in agreement with previous studies based on simulations
(e.g., Boylan-Kolchin, Ma, \& Quataert 2006; Hopkins et al. 2009a).
Naab et al. (2009) showed
that minor mergers or the accretion of relatively low-mass satellites may be the main driver for the late 
evolution of sizes of massive early-type galaxies.
Somerville et al. (2008) showed based on a CDM model of disc formation with $M_*>10^{10}$ M$_\odot$ 
that the average size of discs at fixed stellar mass was about 50 percent larger than that at $z\sim3$.
The predicted evolution in the mean size at a fixed stellar mass since $z\sim1$ 
is about $15\sim20$ percent, which is comparable with our observations. 

The newly accreted small galaxies preferentially populate the outer region of massive galaxies.
The small size growth of the present result is plausibly accounted for by minor mergers or the accretion of relatively 
low-mass satellites (e.g., Naab et al. 2009; Hopkins et al. 2009a).
Naab et al. (2009) and Fan et al. (2010) showed the fractional variation of the gravitational radius of the main 
galaxy after $N$ minor-merger events with $R\propto M_*^\alpha$ as
\begin{eqnarray}
\frac{r_f}{r_i}=\Bigl\{\frac{(1+\eta)^2}{1+\eta^{2-\alpha}}\Bigr\} ^N,
\end{eqnarray}
where $r_f$ and $r_i$ are final and initial galaxy sizes, $\eta$ is the fractional ratio of merging galaxy to 
the main galaxy.
It takes $\sim$1 ($\sim$3) minor-merger events with $\alpha=0.15$ to increase 
the radius by $\sim 20$ ($\sim70$) percent for massive galaxies since $z\sim3$, provided that the merger mass ratio is 1:10. 
It would not be unreasonable that massive galaxies experienced such a small number of minor merging
since $z\sim$3 (e.g., Bundy et al. 2009; L\'opez-Sanjuan et al. 2011).

Deep observations will be important in constraining the exact amount (or lack thereof) and distribution of merging 
galaxies, and how galaxies built up with redshift.
In this context, our finding demonstrates that minor mergers in massive system built up an envelope of lower surface density materials. 
Deeper imaging observations in a wider filed with high spatial resolution and consistent analyses from the local universe 
to high redshift will give 
constraints on the compactness and the amount of low-surface brightness material at the outer envelope 
as a function of redshift to improve our understanding of galaxy formation.

\section*{ACKNOWLEDGMENTS}

This work has been supported in part by a Grant-in-Aid for Scientific Research (21244012) 
of the Ministry of Education, Culture, Sports, Science and Technology in Japan.
We thank Ramsey Lundock for careful reading of the manuscript.
MODS catalogue has been accomplished by MOIRCS builders. We owe the present study to their dedicated efforts.

\label{lastpage}


\begin{thebibliography}{99}
\bibitem[Akiyama et al. (2008)]{akiyama08} Akiyama M., Minowa Y., Kobayashi N., Ohta K., Ando M., and Iwata  I. 2008, ApJS., 175, 1
\bibitem[Barden et al. (2005)]{barden05} Barden A., et al. 2005, ApJ, 635, 959
\bibitem[Bernardi et al. (2011)]{bernardi11} Bernardi M., Roche N., Shankar F., Steth, R.~K., 2011, MNRAS, 412, L6
\bibitem[Bertin \& Arnouts (1996)]{bertin96} Bertin E., Arnouts S. 1996, A\&AS, 117, 393
\bibitem[Bezanson et al (2009)]{bezanson09} Bezanson R., van Dokkum P.~G., Tal T., Marchesini D., Kriek M., Franx M., Coppi P., ApJ., 697, 1290
\bibitem[Bond et al. (2011)]{bond11} Bond N.~A., Gawiser E., Koekemoer A.~M., 2011, ApJ, 729, 48
\bibitem[Bouwens et al. (2004)]{bouwence04} Bouwens R.~J.,  Illingworth G.~D., Blakeslee J.~P., Broadhurst T.~J., Franx M., 2004, ApJ, 611, L1
\bibitem[Boylan-Kolchin et al. (2006)]{boylan06} Boylan-Kolchin M., Ma C-P., Quataert E., 2006, MNRAS, 369, 1081
\bibitem[Bruzual \& Charlot(2003)]{bru03} Bruzual G., Charlot S. 2003, MNRAS, 344, 1000
\bibitem[Buitrago et al. (2008)]{buitrago08} Buitrago F., Trujillo I., Conselice C.~J., Bouwens R.~J., Dickinson M., Yan H., 2008, ApJ, 687, L61
\bibitem[Bundy et al. (2009)]{bundy09} Bundy K., Fukugita M., Ellis R.~S., Targett T.~A., Belli S., Kodama T., ApJ, 697, 1369
\bibitem[Carrasco et al. (2010)]{carrasco10} Carrasco E.~R., Conselice C.~J., Trujillo I., 2010, MNRAS, 405, 2253
\bibitem[Cassata et al. (2010)]{cassata10} Cassata P., et al., 2010, ApJ, 714, L79
\bibitem[Cassata et al. (2011)]{cassata11} Cassata P., et al., 2011, ApJ, 743, 96
\bibitem[Cimatti et al. (2008)]{cimatti08} Cimatti A., et al. 2008, A\&A, 482, 21
\bibitem[Daddi et al. (2005)]{daddi05} Daddi E., et al. 2005, ApJ, 626, 680
\bibitem[Damjanov et al. (2009)]{damjanov09} Damjanov I., et al., 2009, ApJ, 695, 101
\bibitem[Fan et al. (2008)]{fan08} Fan L., Lapi A., de Zotti G., Danese L, 2008, ApJ, 689, L101
\bibitem[Fan et al. (2010)]{fan10} Fan L., Lapi A., Bressan A., Bernardi M., de Zotti G., Danese L, 2010, ApJ, 718, 1460
\bibitem[Fontana et al. (2009)]{fontana09} Fontana A., et al. 2009, A\&A, 501, 15
\bibitem[Franx et al. (2008)]{franx08} Franx M., van Dokkum P.~G., Schreiber M.~F., Wuyts S., Labb\'e, Toft S., 2008, ApJ, 688, 770
\bibitem[Fukugita et al. (1996)]{fuku96} Fukugita M., Ichikawa T., Gunn J.~E., Doi M., Shimasaku K., Schneider D.~P. 1996, AJ, 111, 1748
\bibitem[Guo \& White (2008)]{guo08} Guo Q. White S.~D.~M., 2008, MNRAS, 384, 2 
\bibitem[Hopkins et al. (2009a)]{hopkins09a} Hopkins P.~F., Hernquist, L., Cox T.~J., Keres, D., Wuyts S., 2009, ApJ, 691, 1424
\bibitem[Hopkins et al. (2009b)]{hopkins09b} Hopkins P.~F., Bundy K., Murray N., Quataert E., Lauer T.~R., and Ma, C-P., 2009, MNRAS, 398, 898
\bibitem[Hopkins et al. (20010)]{hopkins10} Hopkins P.~F., Bundy K., Hernquist L., Wuyts S., Cox T.~J., MNRAS, 401, 1099
\bibitem[Ichikawa et al. (2010)]{ichikawa10} Ichikawa T., Kajisawa M., Yamada T., Akiyama M., Yoshikawa T., Onodera M., Konishi M., 2010, ApJ., 709, 741
\bibitem[Kajisawa et al. (2009)]{kajisawa09} Kajisawa M., et al., 2009, ApJ, 702, 1393
\bibitem[Kajisawa et al. (2011)]{kajisawa09} Kajisawa M., et al., 2011, PASJ, 63, S379
\bibitem[Kajisawa \& Yamada (2001)]{kajisawa01} Kajisawa M., Yamada T., 2001, PASJ, 53, 833
\bibitem[Konishi et al. (2011)]{konishi11} Konishi M., et al., 2011, PASJ, 63, S363
\bibitem[Law et al. (2012)]{law12} Law, D.~R., Steidel, D.~C., Shapley, A.~E., Nagy, S.~R., Reddy, N.~A., Erb, D.~K., 2012, ApJ, 745, 85
\bibitem[L\'opez-Sanjuan et al. (2011)]{sanjuan11} L\'opez-Sanjuan C., et al. 2011, A\&A, 530, A20
\bibitem[MacArthur \& Courteau (2003)]{macarthur03} MacArthur L.~A., Courteau S., 2003, ApJ, 582, 689.
\bibitem[Mancini et al. (2010)]{mancini10} Mancini C., et al., 2010, MNRAS, 401, 933
\bibitem[McIntosh et al. (2005)]{mcintosh05} McIntosh D., et al., 2005, ApJ, 632, 191
\bibitem[Mosleh et al. (2011)] {mosleh11} Mosleh M., Williams R.~J., Franx M., Kriek, M., 2011, ApJ, 727, 5
\bibitem[Naab et al. (2009)]{naab09} Naab T., Johansson P.~H., Ostriker J.~P., 2009, ApJ, 699, L178
\bibitem[Nagy et al. (2011)]{nagy11} Nagy S.~R., Law D.~R., Shapley A.~E., Steidel C.~C., 2011, ApJ, 735, 19L
\bibitem[Oke \& Gunn(1983)]{oke83} Oke J.~B., Gunn J.~E. 1983, ApJ, 266, 713
\bibitem[Saracco et al. (2009)]{saracco09} Saracco P., Longhetti M., Andreon S., 2011, MNRAS, 392, 718
\bibitem[Saracco et al. (2011)]{saracco11} Saracco P., Longhetti M., Gargiulo A., 2011, MNRAS, 412, 2707
\bibitem[S\'ersic (1968)]{sersic68} S\'ersic J.-L. 1968, Atlas de Galaxias Australes (Cordoba: Obs. Astron.)
\bibitem[Somerville et al. (2008)]{somerville08} Somerville R.~S., et al. 2008, ApJ, 672, 776
\bibitem[Szomoru et al. (2011)]{szomoru10} Szomoru D., 2010, ApJ.,714, L244
\bibitem[Stott et al. (2011)]{stott11} Stott J.~P., Collins C.~A., Burke C., Hamilton-Morris V., Smith G.~P., 2011, MNRAS, 414, 445
\bibitem[Shen et al. (2003)]{shen03} Shen S., Mo H.~J., White S.~D.~M., Blanton M.~R., Kauffmann G., Voges W., Brinkmann J., Csabai I., 2003, MNRAS, 343, 978
\bibitem[Stockton et al. (2011)]{stockton10} Stockton A., Shih H-Y., Larson K., 2010, ApJ, 709, L58
\bibitem[Suzuki et al. (2008)]{suzuki08} Suzuki R., et al. 2008, PASJ, 60, 1347
\bibitem[Toft et al. (2007)]{toft07} Toft S., et al., 2007, ApJ, 671, 285
\bibitem[Trujillo et at. (2006)]{trujillo06} Trujillo I., et al. 2006, ApJ, 650, 18
\bibitem[Trujillo et at. (2007)]{trujillo07} Trujillo I., Conselice C.~J., Bundy K., Cooper M.~C., Eisenhardt P., Ellis R.~S. 2007, MNRAS, 382, 109
\bibitem[Trujillo et at. (2009)]{trujillo09} Trujillo I., Cenarro A.~J.,  de Lorenzo-C\'aceres A., Vazdekis A.,, de la Rosa I.~G., Cava A., ApJ., 692, L118
\bibitem[Trujillo et at. (2011)]{trujillo11} Trujillo I., Ferreras I., de la Rosa I.~G., 2011, MNRAS, 415, 3903
\bibitem[Valentinuzzi et al. (2010)]{Valentinuzzi10} Valentinuzzi T. et al. 2010, ApJ, 712, 1232
\bibitem[van der Wel et al. (2009)]{wel09} van der Wel A., Bell E.~F., van den Bosch F.~C., Gallazzi A, Rix H-W., 2009, ApJ., 698, 1232
\bibitem[van der Wel et al. (2008)]{wel08} van der Wel A., Holden B.~P., Zirm A.~W., Franx M., Rettura A., Illingworth G.~D., Ford H.~C., 2008, ApJ, 688, 48
\bibitem[van der Wel et al. (2011)]{wel11} van der Wel A., et al., 2011, ApJ, 730, 38
\bibitem[van de Sande et al. (2011)]{sande11} van de Sande J., et al. 2011, ApJ, 736, L9
\bibitem[van Dokkum et al. (2008)]{dokkum08} van Dokkum P.~G., et al., 2008, ApJ, 677, L5
\bibitem[van Dokkum et al. (2009)]{dokkum09} van Dokkum P.~G., Kriek M., Franx M., 2009, Nat, 460, 717
\bibitem[van Dokkum et al. (2010)]{dokkum10} van Dokkum P.~G., et al., 2010, ApJ, 709, 1018
\bibitem[Williams et al. (1996)]{wil96} Williams R.~E., et al. 1996, AJ, 112, 1335
\bibitem[Williams et al. (2010)]{wil10} Williams R.~J., Quadri R.~F., Franx M., van Dokkum P., Toft S., Kriek M., Labb\'e I., 2010, ApJ, 713, 738
\bibitem[Yamada et al. (2009)]{yamada09} Yamada T., et al., 2009, ApJ, 699, 1354
\bibitem[Zirm et al. (2007)]{zirm07} Zirm A.~W., et al., 2007, ApJ, 656, 66
\end{thebibliography}
\end{document}